\documentclass[%
 reprint,
 amsmath,amssymb,
 aps,
]{revtex4-2}

\usepackage{graphicx}
\usepackage{dcolumn}
\usepackage{bm}

\begin{document}

\preprint{APS/123-QED}

\title{A Dough-Like Model for Understanding Double-Slit Phenomena}

\author{%
Ping-Rui Tsai$^{1}$ and 
Tzay-Ming Hong$^{1\ast}$
}

\affiliation{%
$^1$Department of Physics, National Tsing Hua University, Hsinchu 30013, Taiwan, R.O.C.}


\date{\today}

\begin{abstract}
The probabilistic interference fringes observed in the double-slit experiment vividly demonstrate the quantum superposition principle, yet they also highlight a fundamental conceptual challenge: the relationship between a system before and after the measurement. According to Copenhagen interpretation, an unobserved quantum system evolves continuously based on the Schrödinger equation, whereas observation induces an instantaneous collapse of the wave function to an eigen-state. This contrast between continuous evolution and sudden collapse renders the single-particle behavior particularly enigmatic— especially given that quantum mechanics itself is constructed upon the statistical behavior of ensembles rather than individual entities. In this study, we introduce a Double-Slit Diffraction Surrogate Model (DSM) based on deep learning, designed to capture the mapping between wave functions and probability distributions. 
The DSM explores multiple potential propagation paths and adaptively selects optimal transmission channels using gradient descent, forming a backbone for the information through the network. By comparing the interpretability of paths and interference, we propose an intuitive physical analogy: the particle behaves like a stretchable dough, extending across both slits, reconnecting after transmission, allowing detachment before the barrier.
Monte Carlo simulations confirm that this framework can naturally reproduce the characteristic interference and diffraction probability patterns. Our approach offers a novel, physically interpretable perspective on quantum superposition and measurement-induced collapse. 
The dough analogy is expected to extend to other quantum phenomena. In the final, we provide a dough-based picture, attempting to unify interference, entanglement, and tunneling as manifestations of the same underlying phenomenon.

\end{abstract}

\maketitle

\section{Introduction}


In the summer of 1925, Werner Heisenberg laid the groundwork for matrix mechanics\cite{felippa2001historical} on Helgoland Island, Germany. Around the same time, Erwin Schrödinger formulated wave mechanics using a distinct mathematical framework, i.e., differential operators and wave functions \cite{levada2018review}. 
Both formalisms established the foundation for quantum mechanics. The year 2025 marks the centennial of quantum mechanics and has been designated the International Year of Quantum Science. Among its most iconic demonstrations is the double-slit experiment, renowned for revealing path superposition in quantum systems  \cite{bouwmeester2000physics}. However, efforts to determine which path a single particle takes — the so-called “which-way” experiment - are fundamentally constrained by the uncertainty principle since any attempt to obtain path information inevitably destroys the interference pattern  \cite{busch2007heisenberg,durr1998origin,walborn2002double}. As a result, the true nature of the behavior of a single-particle remains an  enduring and deep mystery of quantum physics  \cite{gibbins1987particles,gao2011meaning,goldstein2013reality,ballentine1970statistical}.

This long-missing piece of the puzzle was eventually accepted, almost as if by necessity, as the first principle that nature is fundamentally governed by probabilistic laws \cite{schlegel1970statistical}. Despite persistent doubts regarding the nature of the wave function and the notion of a probability wave, the physics community largely adopted the pragmatic position of “Shut up and calculate.” \cite{kaiser2014history} Proponents regarded further inquiry into such foundational questions as a category mistake \cite{adlam2025kind,bell1990against,einstein1935can}, while critics argued that the superposition merely serves as a linguistic placeholder for phenomena we do not yet understand \cite{albert2009quantum}. Uncomfortable with the idea that events at the quantum level occurred by chance, Einstein’s realist critique that “God does not play dice with the universe” \cite{natarajan2008einstein} ultimately faded from mainstream attention, yet several pivotal follow-up experiments emerged. In spite of the extreme care taken to minimize disturbances to the motion of the photon, the interference pattern was found to vanish, demonstrating that the mere existence of which-way information destroys interference \cite{kwiat1994experimental,robens2018atomic,wiseman1996quantum,svensson2013pedagogical,matzkin2015weak,foo2022relativistic}.
 
The Elitzur–Vaidman bomb experiment further illustrated the relationship between measurement and interference by defining it in terms of whether a bomb explodes — thereby introducing the concept of interaction-free (null) measurement \cite{kwiat1994experimental}. Moreover, Howard Wiseman proposed the theory of weak measurement, employing a method with intrinsic uncertainty that does not destroy the interference pattern while averaging over many trajectories \cite{wiseman1996quantum,svensson2013pedagogical}. Remarkably, the results produced trajectories resembling those predicted by the Bohmian mechanics, offering new insights into the elusive behavior of quantum particles \cite{matzkin2015weak,mahler2016experimental,foo2022relativistic,foo2022relativistic}.

The Copenhagen interpretation has not only been widely adopted to successfully explain quantum phenomena such as the elusive quantum entanglement~ \cite{werner2001bell} and tunneling effects~ \cite{anderson1963probable}, but also applied in practical fields like quantum computing and quantum networks~\cite{sood2023quantum,gyongyosi2019survey,wei2022towards,huang2025decoherence,ho2024online}. Nonetheless, there exists several alternative interpretations, including the de Broglie–Bohm theory which involves nonlocal hidden variables and does not require wave-function collapse\cite{bohm1961hidden}, the interpretation of many-worlds\cite{albert1988interpreting}, and quantum Bayesian theory~\cite{holland1993broglie,plaga1997possibility,timpson2008quantum}. Collectively, these perspectives have enriched and expanded the conceptual landscape for interpreting quantum mechanics.


\section{Double-slit experiments}
Several experimental variations and conceptual extensions of the double-slit experiment have been investigated. Noteworthy examples include:



(1) In 1959, Aharonov and Bohm (AB) proposed the concept that even in regions where both the electric and magnetic fields are zero, the vector potential $\mathbf{A}$ and scalar potential $U$ can still influence quantum particles, producing observable effects such as phase shifts in a double-slit experiment  \cite{aharonov1959significance}. This suggests that these potentials may represent more fundamental physical quantities. The AB effect was later experimentally confirmed by Tonomura {\it et al.}\cite{tonomura1986evidence} and Osakabe {\it et al.}~\cite{osakabe1986experimental} in 1986, demonstrating the gauge-invariant geometric phase~\cite{shapiro1989aharonov} and the nonlocal nature of quantum systems.

(2) In 1999, C$_{60}$ molecules were employed by Arndt {\it et al.}  \cite{arndt1999wave} to demonstrate that interference fringes persist even at the mesoscopic scales, thereby providing strong evidence that wave-like behavior is not confined to the microscopic domain.

(3) In 2017, Maga\~{n}a-Loaiza {\it et al.} showed that, by shining a laser on a surface engineered with surface plasmon, photons can bend and sequentially traverse multiple slits, transforming the conventional double-slit diffraction pattern into a triple-slit interference pattern  \cite{magana2016exotic}. Their results suggest that the principle of superposition should be regarded as an approximate description of a more intricate physical reality~ \cite{sorkin1994quantum,sinha2010ruling}.

Most recently in 2025, (4) Celso J. Villas-Boas {\it et al.}  \cite{villas2025bright} further proposed that the bright and dark states of light collectively form the constructive and destructive interference fringes. In addition, (5) Fedoseev  {\it et al.} \cite{fedoseev2025coherent} investigated how the slit characteristics influences the interference formation. Their meticulous high-precision laser experiments with ultracold atoms beautifully demonstrates that positional uncertainty in atomic lattices plays a crucial role in sustaining interference.

Since the advent of quantum mechanics 125 years ago with Max Planck's proposal of energy quanta to explain black-body radiation \cite{planck1978gesetz}, it has continued to advance; for instance, the successful realization of single-photon sources \cite{lounis2005single} and the true imaging of single photons \cite{yuen2024exact} that were once deemed impossible by early quantum physicists.
Notwithstanding, fundamental problems in the interpretation of double-slit experiments remain, such as why a quantum particle passing through the slits seems to exhibit multiple “copies” that propagate simultaneously in a probabilistic manner \cite{dirac1981principles}.
Moreover, the relationship between a system before and after measurement can appear as if the propagation of wave function becomes discontinuous as it instantaneously collapses into a definite position, effectively ignoring the intermediate processes \cite{teller1983projection, margenau1936quantum, bell2004speakable, cushing1994quantum}.
Although the principle of complementarity attempted to rationalize this peculiar behavior, recent surveys indicate that only 40$\%$ of scientists fully embrace the Copenhagen interpretation \cite{gibney2025physicists}.
The concept of the “black box” arises from the fundamental differences in how physical quantities are interpreted in classical and quantum measurements \cite{landsman2006between,baggott2004beyond}.

In classical mechanics, there exists a clear causal relationship between an object and the measuring apparatus. Physical quantities are assumed to preexist prior to observation, and measurement merely reveals these values without fundamentally disturbing the system.
In contrast, quantum mechanics completely overturns this perspective.
The Schrödinger equation describes a continuous, statistical evolution of the wave function, which does not directly correspond to observable physical quantities, nor depict definite trajectories over time. Instead, it provides the probability distribution of possible outcomes \cite{berezin2012schrodinger}.
Although this evolution appears continuous and causally structured, it does not establish a deterministic link between measurement results at different times \cite{schrodinger1935present,bell2004speakable, von2018mathematical}.

Consequently, the internal reality of a quantum system is concealed within an inaccessible “black box”.
We can only construct a theoretical framework using operators and wave functions, which lack direct physical meaning.
The intermediate evolution during measurement is neglected and summarized as an instantaneous state determination — the so-called “wave function collapse”\cite{von2013mathematische}.
This instantaneous collapse introduces unavoidable conceptual challenges.
In particular, the transition from a probabilistic superposition to a definite outcome cannot be understood by classical intuition, nor can the interactions between the measured system and the measuring apparatus be easily separated and/or interpreted.

As an interpretation of quantum measurement, the wave function represents a compromise between realism and the intrinsic nonlocality of the measurement process.
Because the projection postulate implies a form of nonlocal state update\cite{aerts2013quantum}, various attempts have been made to reconcile these results using local hidden-variable theories\cite{einstein1935can}.
However, such theories have been ruled out by their violation of the Bell’s inequalities\cite{peres1999all,blaylock2010epr} and related experimental findings, and cannot reproduce the statistical predictions of quantum mechanics.

\begin{figure}
\centering
\includegraphics[width=8.5cm]{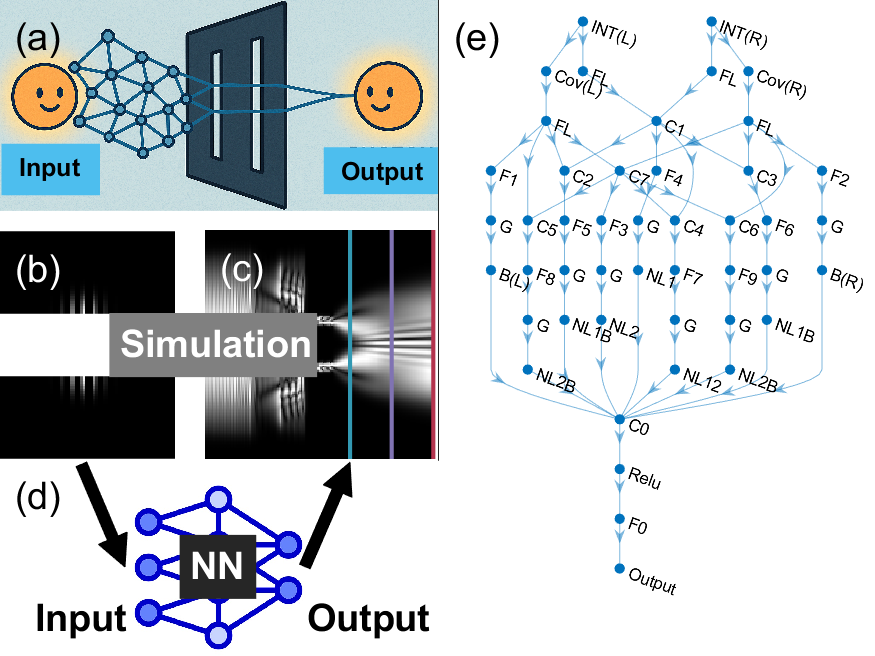}\caption{Schematic layout of our approach to revisit the double-slit phenomenon via deep learning. (a) The interference framework is treated as an equivalent model, where input particles produce the output interference fringes. (b) The probability distribution results under the evolution of the input and output wavefunctions obtained via numerical simulation are presented in (c), with blue, purple, and red lines representing positions -  just exiting, moderate and long distances from the slits. These results are then fed into the neural network depicted in (d). (e) By utilizing the latent path information within the DSM and applying gradient descent, the network selects the optimal paths for the interference probability distributions at different times, thereby providing possible insights for the theoretical representation.}
\end{figure}

\section{Studying double-slit phenomenon by deep learning}
If experiments alone cannot serve as an effective method, can we infer or construct a model that “substitutes” for the double-slit diffraction process? If such a model can be established, analyzing the internal information of the black box may provide insights that circumvent experimental limitations\cite{busch2007heisenberg}. This idea aligns with developments in engineering and deep learning\cite{lecun2015deep}, where global surrogate models have been used to efficiently approximate complex systems\cite{liu2021global, cheng2020surrogate}. Inspired by these advances, constructing a surrogate model for double-slit diffraction may offer a new pathway to overcome experimental challenges. To implement this idea, we adopt an input–output signal-processing framework, modeling double-slit diffraction as the system depicted in Fig. 1(a). We anticipate that this approach could provide new insights into the principle of complementarity and the dynamic processes of quantum behavior behind the double slit.

The two-dimensional complex quantum wave packets obtained from numerical simulations serve as the input, while the corresponding interference fringe patterns form the output database for a deep convolutional neural network (NN) \cite{li2021survey,guo2020first}, as illustrated in Fig. 1(b$\sim$d). Convolution, as a fundamental mathematical operation in signal processing\cite{krishna2017digital}, enables the network to learn transformations\cite{han2022survey,li2021survey}; training by gradient descent is analogous to the principle of least action \cite{guo2020first}. Moreover, in the context of signal processing, convolution also represents the concept of impulse response of a system.

Although deep learning may appear superficial as another "black box", numerous interpretability methods now exist to analyze the contribution of hidden layers and inter-layer features to predictive performance\cite{zhang2018visual,chakraborty2017interpretability}. Such approaches have been widely applied in brain–machine interfaces and neuroscience\cite{eickenberg2017seeing,ferrante2025towards}, for example, by modeling the correspondence between perceptual processing and deep learning to understand the temporal dynamics of visual processing\cite{eickenberg2017seeing}. If we can establish a  Diffraction Surrogate Model (DSM) for the double-slit experiment that takes quantum wave packets as input and outputs interference probability distributions, then decomposing how the wave-function information is processed in space and time may provide a novel avenue for breakthroughs in understanding quantum dynamics.

The application of artificial neural networks (ANN) to studying physical systems is not new. For example, this  concept has been implemented in Neural Quantum States (NQS) approach for many-body quantum problems by representing the wave functions within neural architectures and optimizing their energy through variational methods\cite{jia2019quantum, reh2023optimizing}. Similarly, physics-inspired deep learning has also sought to incorporate the governing principles of physics directly into neural networks\, enabling them to learn, simulate, and control complex physical phenomena through an artificial intelligence framework\cite{cuomo2022scientific, cai2021physics}.

\begin{figure}
\centering
\includegraphics[width=8.5cm]{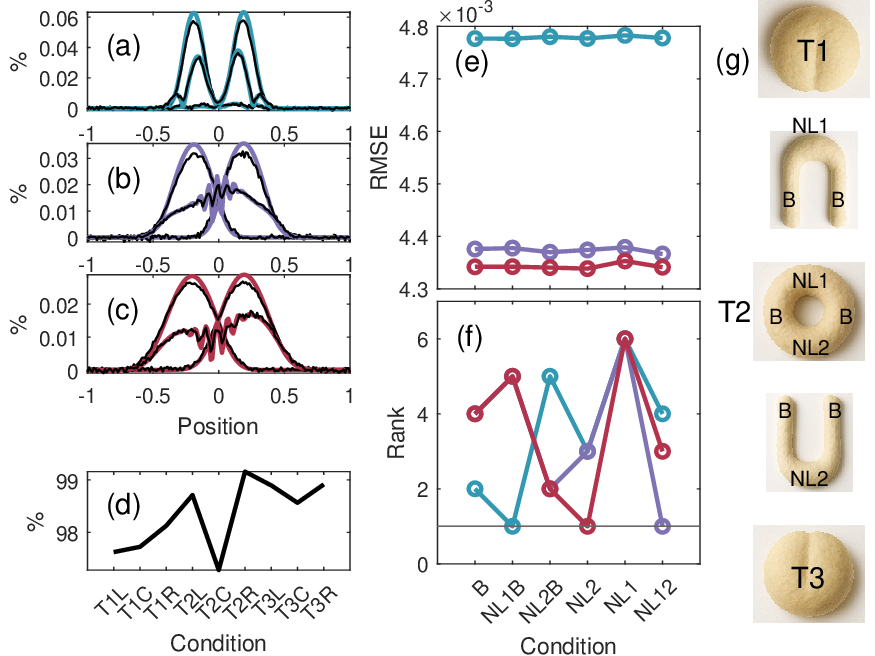}\caption{The color scheme follows that of Fig. 1(c). (a–c) correspond to the probability distributions of the three colored lines in Fig. 1(c), with each subplot containing three samples representing the interference when the wavefunction enters the slit slightly left, centered, or slightly right. The black line shows the predicted outcome by the DSM. (d) shows the similarity between the probability distributions from the numerical simulation and the DSM predictions. (e) compares the RMSE of the encoding models of three DSM models under different EL values. (f) ranks the models in (e) from smallest to largest RMSE. (g) illustrates the possible dynamic processes of particles passing through the double slits based on the ranking in (f), where the red and blue points represent NL1 and NL2.}
\end{figure}

Our objective is to address two main questions:
(Q1) How can the DSM inspire the construction of a physical picture after the double slit?
(Q2) How can the physical picture derived from the DSM be used to construct a complete double-slit interference simulation from scratch?
To answer them, we designed three works.
(W1) To analyze the backbone network of the DSM to understand the temporal usage of paths by quantum information, we adopt an encoding model commonly used in neuroscience and deep learning. By evaluating the interpretability between the feature maps of individual paths within the DSM and the final interference probabilities, we  infer the patterns potentially employed by the model during information processing. As shown in Fig. 1(e), the NN structure in our case relies on the gradient descent training procedure in deep learning, which drives the model to select paths that minimize error and achieve optimal performance. Based on this, we constructed a linear regression model between the input feature maps and the final interference probabilities for each path within the DSM, and used RMSE to assess interpretability rankings to determine the backbone network\cite{simas2021distance,serrano2009extracting}. By comparing the networks at three different time points in the DSM, we analyzed the differences in main paths across temporal stages. 
Denoted by time T1, T2, and T3, Fig. 2(a$\sim$c) correspond  respectively to when interference has not yet, just slightly, and fully developed. 

(W2) To analyze interference and non-interference features in the two-dimensional interference distribution, we use simulations of the two-dimensional double-slit diffraction probability distributions as a database to investigate the effect of disrupting latent variables in the network and observe the decoder reconstruction results. The t-SNE is then applied for clustering to interpret the visual meaning of these patterns.

(W3) To construct a physical model, we propose a physical model, based on the clues from (W1) and (W2), and employ a Monte Carlo method to reproduce the full double-slit interference probability distribution. In summary, tasks (W1) and (W2) correspond to (Q1), while (W3) address (Q2). 

\section{Details of simulation procedures and results} 

The arrangement of this section includes: the construction of the database required to generate the DSM in Sec. IV(A), the outcome of (W1) in Sec. IV(B), additional analyzes of the database for (W2) in Sec. IV(C), and   the results from (A$\sim$C) are eventually integrated to develop a theoretical framework,  
and Monte Carlo simulations is employed to model the double-slit experiment for (W3)  in the final Sec. IV(D).

 
\subsection{Preparing the database of double slits}

The training data presented in Fig. 1 were generated through numerical simulations of the time-dependent Schrödinger equation within a two-dimensional spatial domain incorporating a double-slit potential barrier\cite{mena2021solving,artmenlope2021double-slit}. The governing equation is
\begin{equation}
i\hbar\frac{\partial \psi(x,y,t)}{\partial t}
= \left[-\frac{\hbar^2}{2m}\nabla^2 + V(x,y)\right]\psi(x,y,t),
\end{equation}
where $V(x,y)$ denotes the double-slit potential used in the simulations.

Each sample is initialized as a Gaussian wave packet with a plane-wave phase:
\begin{equation}
\psi(x,y,0) = \exp\!\Big[-\frac{(x-x_0)^2}{2\sigma_x}-\frac{(y-y_0)^2}{2\sigma_y}\Big]\,
\exp\big(i k (x-x_0)\big),
\end{equation}
where $k$ is the central wavenumber and $y_0$ denotes the initial $y$-center (set proportional to $x_0$ in the code).

The parameter ranges used to generate the dataset of 3,200 samples are summarized as follows:
\begin{itemize}
  \item initial $x$-center: $x_0 \approx 0.40$ to $0.60$ (21 discrete values);
  \item horizontal width: $\sigma_x$ sampled over a set of small dispersions (38 discrete values);
  \item vertical width: $\sigma_y$ sampled over a subset of small dispersions (5 discrete values).
\end{itemize}

The wave function is evolved in time on a square domain of side length $L$ with hard-wall boundary conditions ($\psi=0$ at the edges). Time evolution is performed numerically (implicit time-stepping) and the modulus $|\psi(x,y,t)|$ is recorded at each time step for each parameter combination.

\subsection{Analyzing the backbone network of DSM}

In the neural network (NN) architecture, the model itself does not inherently possess a notion of physical space. Therefore, in our designed DSM, the simulated incident wave function is preprocessed by dividing it into left and right components, denoted as \text{INT(L)} and \text{INT(R)} in Fig.~1(e), both resolution are 33 $\times$ 70. In the following descriptions, the “/” symbol is also used to indicate the left and right flankers. The number of samples is 3200, and the training:validation:testing ratio is 8:1:1, overfitting is not a major concern here, because the DSM merely relies on a neural network to model the relationship between inputs and outputs.

Each of the INT inputs has a branch directed toward a convolutional layer (\text{Cov}) \cite{o2015introduction} to extract spatial features corresponding to the propagation through a local slit:
\begin{align}
\text{INT(L)} &\rightarrow \text{Cov(L)} \rightarrow \text{FL} \rightarrow \text{F1}\rightarrow \text{G}\rightarrow \text{B(L)}\\
\text{INT(R)} &\rightarrow \text{Cov(R)} \rightarrow \text{FL} \rightarrow \text{F2}\rightarrow \text{G}\rightarrow \text{B(R)}
\end{align}
Here, each convolutional layer employed a kernel size of $4 \times 4$, a stride of $1 \times 1$, and same padding with a zero-padding value where \text{F} denotes the fully connected layer. The GELU activation function \text{G} serves to prevent potential cases in which a specific pathway carries no informational flow. The $\text{B(L/R)}$ represents one of the final Flatten layers ($\text{FL}$) in the network, describing the contribution corresponding to a single slit. In our model, each branch's $\text{FL}$ has a dimension of 128.


Because this left–right separation is artificial, the additional pathway represents the integration of information between the two inputs prior to their passage through the slits, implying a possible non-local relationship. To achieve this, both input branches are passed through (\text{FL}) and subsequently concatenated (\text{C}) to form a unified representation of the interference response:
\begin{align}
\text{INT(L)} &\rightarrow \text{FL}, \\
\text{INT(R)} &\rightarrow \text{FL}, \\
C[(5), (6)] &\rightarrow \text{C1}.
\end{align}
The concatenated vector \text{C1} is then passed through \text{F4}, followed by \text{G}, yielding the branch output referred to as the non-local term named \text{NL1}. The subscript “1” indicates that this term represents the correlations between the two inputs prior to their passage through the slits.

We also consider the combined operation between \text{B} and \text{NL1}, representing the simultaneous contribution of both local and non-local information. This branch is defined as follows:
\begin{align}
\text{Cov(L/R)} &\rightarrow \text{FL}, \\
C[(8), \text{C1}] &\rightarrow \text{C2/C3} 
                 \rightarrow \text{F5/F6} 
                 \notag\\
                 &\rightarrow \text{G} 
                 \rightarrow \text{NL1B(L/R)}.
\end{align}
The second non-local term, denoted as \text{NL2}, is hypothesized to arise from the recombination of information flows from the two \text{Cov(L/R)} layers. Here, the subscript “2” indicates that this process occurs after the passage through the slits, as expressed by:
\begin{align}
\text{INT(L)} &\rightarrow \text{Cov(L)} \rightarrow \text{FL}, \\
\text{INT(R)} &\rightarrow \text{Cov(R)} \rightarrow \text{FL}, \\
C[(10), (11)] &\rightarrow \text{C7} \rightarrow \text{F3} \rightarrow \text{G} \rightarrow \text{NL2}.
\end{align}

Similar to \text{NL1B}, we also consider the pathway combining \text{NL2} and \text{B}. 
This branch is defined as follows:
\begin{align}
C[(8), \text{C7}] &\rightarrow \text{C5/6} \rightarrow \text{F8/9} \rightarrow \text{G} \rightarrow \text{NL2B(L/R)}.
\end{align}
Finally, all FL of the paths, including \text{B(L/R)}, \text{NL1}, \text{NL1B(L/R)}, \text{NL2}, \text{NL2B(L/R)}, and \text{NL12}, 
are concatenated into \text{C0}, followed by a ReLU activation and the final fully connected layer \text{F0}, 
producing the final probability distribution with a resolution of 165. 

Based on the above results, we could construct a conceptual DSM that includes multiple possible pathways. Based on gradient descent, DSM determines which path is most effective for reconstructing the probability distribution. We further examine the internal computations of the DSM using interpretability models to identify the main pathways. After all these efforts, it is high time to draw inspirations on the  quantum information behind the double slit processes by studying DSM . 

To achieve this, we employed the Deep Learning Toolbox - MATLAB \cite{matlab2025deeplearning} for training, with a learning rate of 0.004, MaxEpochs of 2, MiniBatchSize of 8, and an output interference fringe resolution of 79, as shown in Fig. 2(a$\sim$c). The resolution of the left-right segmented 2D quantum wave function (QWF) in all three DSM is $45\times 79$ and complex-valued. 





After training, we evaluate the predictive performance of the DSM at three different time points, T1$\sim$T3 (depicted in light blue, purple, and burgundy). These correspond to the evolution of the two-dimensional dynamic probability amplitude: immediately after emerging from the slits without any interaction, the moment the amplitudes first interact and interference begins, and the final stage showing the fully developed interference pattern. In Fig. 2(d), the  T in the X-axis represents the time points, while LCR indicates whether the wavefunction is located in the left, center, or right region of the slit. To quantitatively compare the DSM's predicted distribution ($P_{\rm DSM}$) with the numerically simulated distribution ($P_S$), we calculate the similarity \cite{d2020two}  by use of the formula:
\begin{equation}
S = \frac{\sum_m \sqrt{|P_{\rm DSM}(m)| \cdot |P_S(m)|}}{0.5 \times \left(\sum_m |P_{\rm DSM}(m)| + \sum_m |P_S(m)|\right)}
\end{equation}
where $m$ denotes the positions of the interference pattern.

Since both the neural network prediction and the numerical distribution may contain small negative values due to noise or background subtraction, we first apply an element-wise absolute value to both $P_{\rm DSM}$ and $P_S$ to ensure non-negativity. The similarity $S$ thus represents the normalized overlap between the two distributions and is expressed as a percentage by multiplying by 100. This measure provides a robust quantification of how closely the DSM follows the expected numerical behavior.

Next, we analyzed each FL separately by computing the linear regression RMSE at each position of the final probability distribution and averaging the results, yielding Fig.~2(e). This analysis helps us understand how the DSM selects the optimal trajectory to achieve the final outcome during the evolution of the probability distributions from T1 to T3. 
From the ranking of FL contributions based on RMSE in Fig. 2(f), we found that NL1B dominates the rank 1 pathway at time point T1, NL12 at T2, and NL2 at T3. 
In the rank 2 pathways, B dominates at T1, indicating that both branches simultaneously contribute to the information flow. At T2 and T3, NL2B becomes dominant, reflecting the cooperative interaction between NL2 and B.

A possible interpretation is as follows: after interacting with the slits, two branches represented by B are formed, and the quantum entity may maintain NL1B, implying that both the influence of the slits on the particle and the connection between the two inputs are simultaneously taken into account. At T2, when interference fringes begin to emerge, both the pre-slit and post-slit non-local connections must be considered concurrently, indicating a continuous link between the two regions and forming a topology NL12 with a single “hole.” By T3, all information is carried primarily by NL2. Nevertheless, at both T2 and T3, the interaction between the slit-related branch B and NL2 remains active. This proposed picture differs from Dirac’s interpretation, in which photons are often described as splitting; In our model, although the information flow allows for recombination and splitting, the quantum entity never actually separates into two independent systems. The conceptual picture derived from the encoding model is presented in Fig. 2(g).

\subsection{Analyzing the interference pattern}

To analyze the interference in more detail, we examined the overall two-dimensional interference distributions under different simulations using a Variational Autoencoder (VAE)\cite{doersch2016tutorial}. We modeled the results  in Fig. 3(a) by using a latent space of 128 dimensions and performed a typical network ablation analysis. By systematically ablating individual points in the latent layer and subtracting the reconstructed images from those without ablation, we obtained the contribution of each latent unit. The t-SNE clustering with 4 clusters is then employed. By averaging the images of each cluster, we obtained the feature maps exhibiting interference fringes in Fig. 3(b$\sim$d), where two distinct classical paths corresponding to direct contributions from the slits are evident. Since t-SNE\cite{maaten2008visualizing} clustering results vary between runs, and given the symmetry of the left and right slits, we selected the t-SNE output with the most symmetric left–right information as the optimal solution. We then compared the grayscale statistics of the slit features using four order parameters: mean, variance, skewness, and kurtosis. After performing 1000 t-SNE runs, we chose the result with the minimal Euclidean distance as the target. This approach is analogous to the Structural Similarity Index (SSIM) \cite{bakurov2022structural} used in image analysis.

\begin{figure}
\centering
\includegraphics[width=8.5cm]{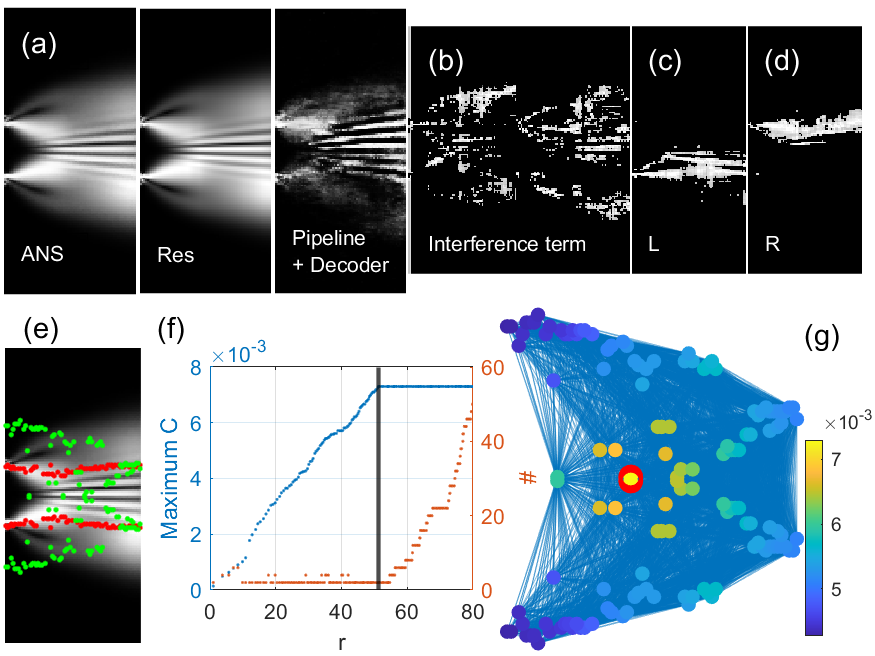}\caption{(a) The double-slit result was obtained from 3,200 simulation samples and with resolution of 165$\times$76. The answers (ANS) represents the ground truth under numerical simulation, also one of the sample input to the VAE. whiles, the
reconstructed image (Res) denotes the reconstructed results after the decoder, and “Pipeline” refers to the images reconstructed by the VAE decoder after all EL values in the DSM are pipeline-transformed into the VAE latent space.
(b$\sim$d) The clustering results of Res obtained by t-SNE with k $=$ 4. Note that (b) represents the two interference terms, while (c) and (d) correspond to non-interference information output from the two slits. (e) The representative information obtained by averaging (b) and (c, d) from left and right and taking the maximum value. (f) The DSM-based hypothesis of a slit convergence point for NL2. We expect this point to be a node with the highest closeness centrality and the fewest quantity. The blue line shows the maximum closeness centrality (MC), the orange line indicates the number of nodes with that MC, $r$ represents the minimum distance connecting different nodes, and the black line indicates the optimal $r$. (g) The network modeling results under the optimal $r$. The color bar is for different degrees of centrality, and the red-circle  indicates the positions of MC.}
\end{figure}  

Based on the results of the encoding model, we hypothesized the existence of a convergence point for NL2. To investigate this, we analyzed the images in Fig. 3(b$\sim$d). To further reduce  the left–right asymmetry, we first averaged the left and right sides, then selected only those features with the maximum grayscale values, as shown in Fig. 3(e) where red dots constitutes the classical paths, and green ones  are from interference terms. We propose that NL2 corresponds to the reintegration of information after having passed the slits, ultimately converging into a single quantum particle. Therefore, we modeled the network based on the green interference points in Fig. 3(e). As shown in Fig. 3(f), the threshold distance $r$ is selected 
to ensures a consistent number of sources and maximum centrality of the closeness\cite{zhang2017degree} for the construction of an optimal network model. These procedures are commonly used in topological data analysis (TDA)\cite{chazal2021introduction}. An initial node located at the overlap region  and closest to the two waves from the slits can then be identified, and highlighted by the red circle, as in Fig. 3(g).

\begin{figure}
\centering
\includegraphics[width=8.0cm]{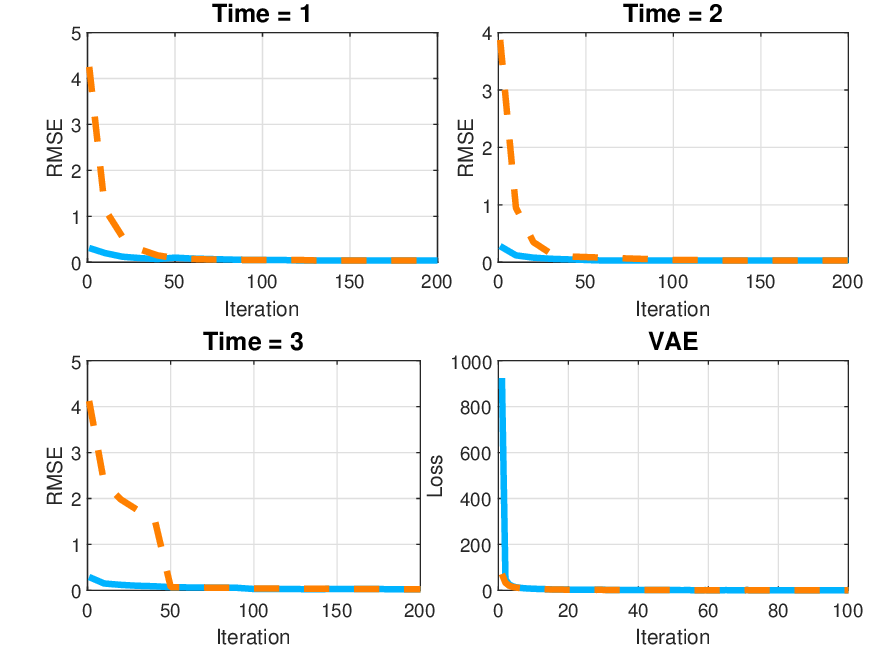}\caption{Learning curves for three DSMs and VAE where the blue/orange lines indicate the training/validarion curvse.}
\end{figure}  

The learning curves of the three DSMs and the VAE, including both training and validation, are shown in Fig. 4. The light bluish-purple and red colors indicated in Fig. 1(c) and Fig. 2(a$\sim$c) are defined as Time = 1$\sim$3. Three testing rate of DSMs are 7.81, 7.95 and 5.77 $\times 10^{-4}$

\subsection{Constructing a physical model}  



In view of the results presented in Sections~IV~A and~B, the DSM and VAE reveal the following insights:
\begin{enumerate}
    \item The Rank~1 result provided by the DSM indicates that the sequence 
    \textbf{NL1B $\rightarrow$ NL12 $\rightarrow$ NL2} represents the operational backbone network of prediction T1$\sim$3 probability. 
    We conceptualize \textbf{NL1B} as reflecting the transmission through the slits and the connection 
    between the two parts of particle, which play equally important roles in T1. 
    During the transmission, \textbf{NL12} encapsulates both slits, and finally, 
    all process information converges into \textbf{NL2}, resembling a deformable dough capable of integrating and reshaping the overall dynamics.

    \item The Rank~2 result from the DSM shows that the difference between 
    \textbf{T1} (without interference fringes) and \textbf{T2}/\textbf{T3} (with fringes) depends on \textbf{NL2B}. 
    This implies that the formation of nonlocal correlations in \textbf{NL2} 
    and the information transmitted through the \textbf{L} and \textbf{R} slits 
    are equally important in the network representation.

    \item The VAE suggests the existence of an origin point at the center between the two slits, that is responsible for producing the interference features. 
When combined with the Rank~1 conclusion of the DSM, it is possible that this origin corresponds to \textbf{NL2}. 
Furthermore, by incorporating the Rank~2 results, both \textbf{NL2} and \textbf{NL2B} must be jointly considered to fully describe the system’s behavior.
\end{enumerate}

The analogy to a dough may sound outlandish. But it offers a tremendous advantage over the black box in the Schrodinger picture since we get to visualize how the particle gets transported through and deformed by the slits. In the following paragraphs, we shall elaborate and build a minimal physical model around this dough scenario to self-consistently explain how the interference pattern, credited to the wave nature and its superposition by Schrodinger, can be realized by random interactions of dough with the slit openings.  

\begin{enumerate}
    \item Before entering the slits, it appears that each quantum particle is randomly assigned a transmission ratio between the two slits. We hypothesize that this randomness originates from the time-dependent positions of atoms or molecules near the slit, whose vibration continuously alters the local spatial configuration. This results in different stretching forces on the quantum particle from the left and right sides of the slit, and different effective transmitted mass through the two slits. Nevertheless, the pre-slit connectivity, denoted as NL1, is preserved, and the particle does not rupture when stretched - hence the likeness to a dough.
    In this model, we employ nine configurations to determine the transmission ratio,
    defined as    
    \begin{align}
        M_L &= M \cdot \frac{W_{L1}}{W_{L1} + W_{R1}},  \\
        M_R &= M \cdot \frac{W_{R1}}{W_{L1} + W_{R1}} 
    \end{align} 
    We list the discrete weights $W_{L1}$ and $W_{R1}$ from 1 to 4, forming 16 possible left-right mass ratio configurations.


\item When passing through the slits, the quantum particle interacts with the atoms or molecules near the slit boundaries. 
The exact physical origin of these forces \(F_L, F_C, F_R\) is not crucial for our later arguments.
Besides the central region (\(F_C\)) between the slits, the quantum particle passing through the left (\(F_L\)) slit interacts with and receives forces from the atoms or molecules near the left edge of the slit. Likewise, the right boundary provides (\(F_R\)) for the passage through the right slit.
Each of \(F_L, F_C, F_R\) is allowed to take one of four discrete values to represent stochastic variations in the set of Monte Carlo events for the slit, 
resulting in \(4^3 = 64\) possible set of Monte Carlo events. 
In each Monte Carlo iteration, one configuration is randomly selected, and the effective forces acting on the particle by the left and right slits are calculated as:
\begin{align}
    F_{\text{left}} &= F_L + F_C, \\
    F_{\text{right}} &= F_R + F_C
\end{align}
that guide the branch trajectories and determine the subsequent evolution of the system.

\item Based on the DSM results, this model correctly predicted two high-probability sites if the screen is immediately behind the slits. 
Since the branch of dough through the left and right slits is not symmetrical and carries different mass, 
we hypothesize that they merge and recombine at the same position as their center-of-mass:
\begin{align}
    Y_C = \frac{M_L Y_L + M_R Y_R}{M_L + M_R},
\end{align}
where $Y_L$ and $Y_R$ represent the slit positions as viewed by the observer.

\item As in classical dynamics, the motion of “dough-like” particle is governed by the net force, $F_{\text{tot}} = F_{\text{left}} + F_{\text{right}}$ that acts on its center-of-mass. 
In our model, the resultant force is simplified to act only along the $Y$-direction,  parallel to the plane of the slits.  
While performing Monte-Carlo, we artificially set an interaction time $t_{I}$ beyond which $F_{\text{tot}}$ is turned off. Note that all references to time correspond to the iteration steps.

4.1. \textbf{Acceleration phase} ($t < t_{I}$): During this phase, the particle follows the constant-acceleration motion:
\begin{align}
    Y(t) = Y_C + \frac{1}{2} \frac{F_{\text{tot}}}{M} t^2
\end{align}
where $Y_C$ is the center-of-mass position prior to the slits, $M$ is the total mass of the particle, and $t$ is the elapsed time.  

4.2. \textbf{Uniform motion phase} ($t \ge t_{I}$): After the critical interaction time, the particle continues to move with the instantaneous velocity attained at $t_{I}$. The position is updated at $t_{I} = 15$ and afterwards follow
\begin{align}
    Y(t) = Y(t_{I}-1) + \frac{t_{I} F_{\text{tot}}}{M} (t - t_{I} + 1),
\end{align}
where $Y(t_{I}-1)$ represents the particle's last position under the force, and the second term corresponds to displacement due to uniform motion with velocity $v = t_{I} F_{\text{tot}} / M$. 
This formulation ensures continuous and physically consistent trajectories from the acceleration to uniform motion  phases. Simulation results are shown in Fig. 5. 
\end{enumerate}

\begin{figure}
\centering
\includegraphics[width=8.5cm]{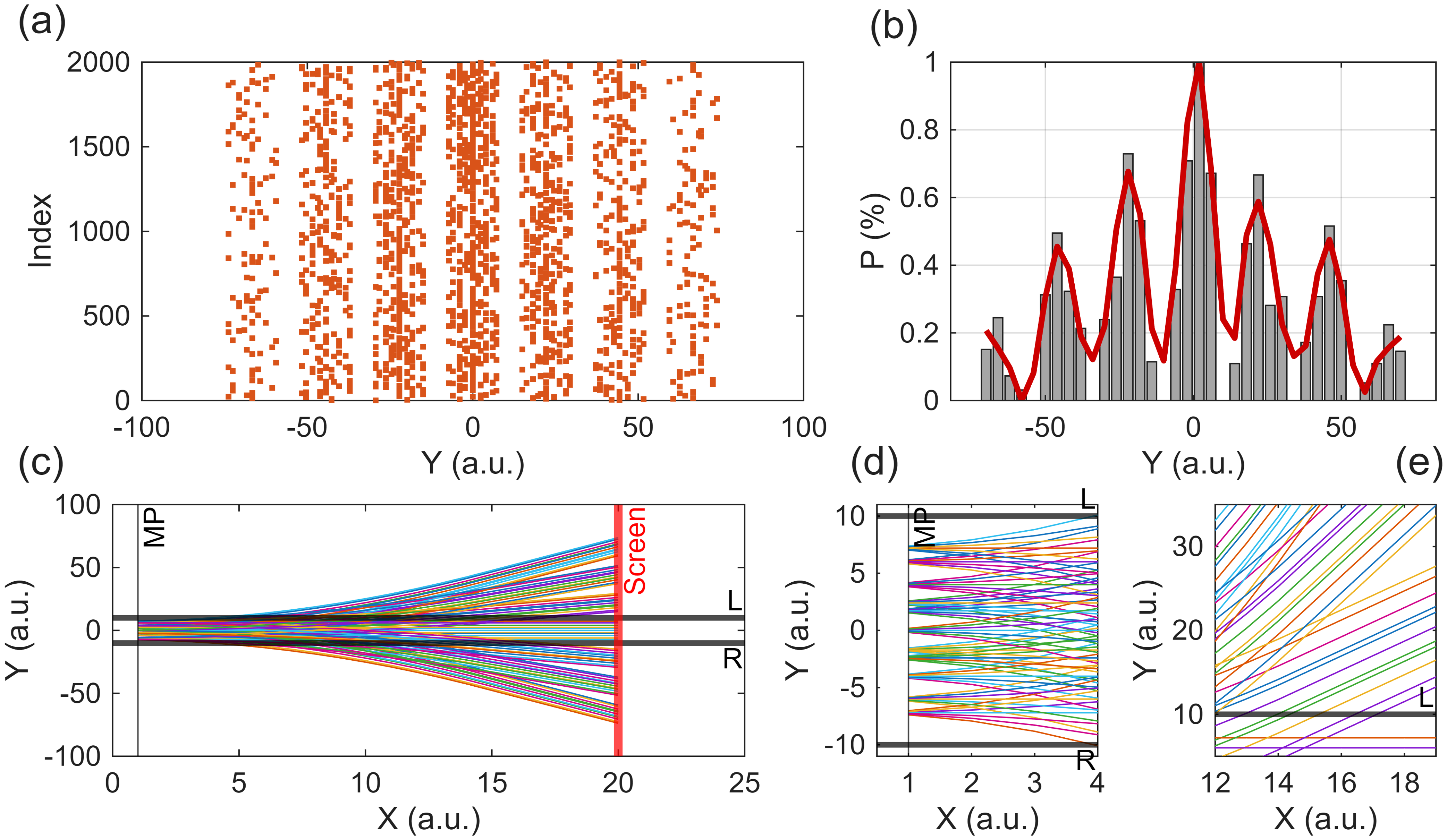}\caption{(a) The spatial distribution of the arrival positions on the screen after 2000 inputs. (b) The probability distribution is obtained by drawing histogram with a bin size of 4 units, where the red curve is to highlight the interference pattern. (c) Trajectories from MP to the screen. (d, e) Bifurcation results near MP and along two selected stripes of trajectories.}
\end{figure}

\begin{figure}
\centering
\includegraphics[width=8.5cm]{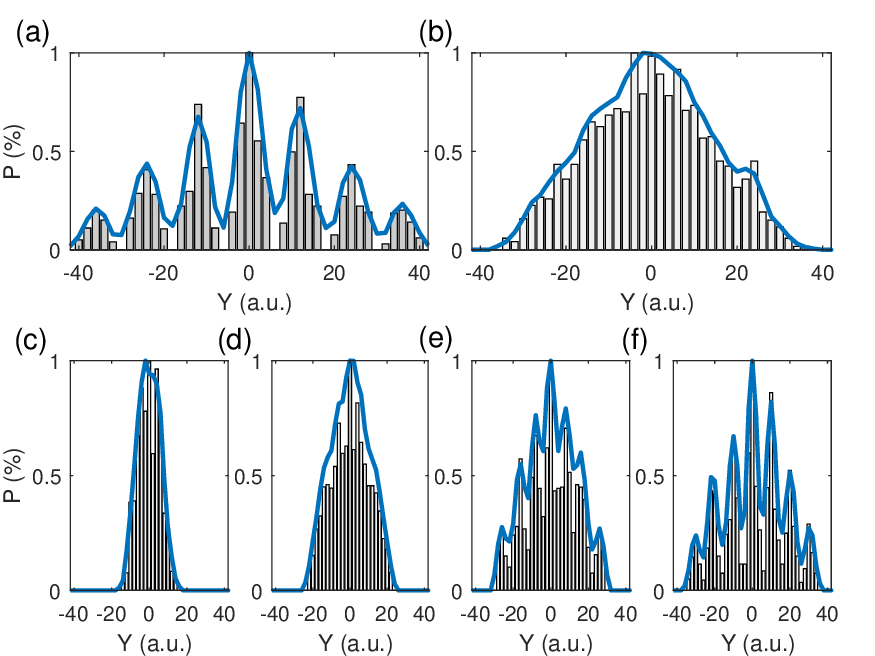}\caption{(a) and (b) show the probability distribution on the screen under two different sets of simulation parameters. 
(c$\sim$f) present the probability distributions for a fixed configuration number of $4^3$ with $t_{I}$= 3, 6, 9, and 12, respectively.
}
\end{figure}  


Figure 5(a) shows the results of 2000 \cite{kolenderski2014time} Monte Carlo simulations \cite{rubinstein2016simulation}, with each marker representing a single Monte Carlo outcome. Under these simulation conditions, we can clearly observe equally spaced bright fringes, reproducing the interference pattern, with the correct decreasing rate in intensity as one moves away from the central fringe. The results from 
Fig. 5(a) are further represented as a histogram in Fig. 5(b). 
Figure 5(c) illustrates the trajectories of particle beyond the merge position (MP) where the two branches form the left (L) and right (R) slits recombined. Figures 5(d) and 5(e) show a magnified view beyond MP and the bifurcation of two adjacent bright fringes, respectively.

To investigate the effect of set of Monte Carlo events and the choice of $t \ge t_{I}$ on the simulation outcomes, we simulate under two parameter sets in Fig. 6(a, b).
The mass ratio between the left and right branches through the double slits, $W_{L1}$ and $W_{R1}$, were chosen to take discrete values of ${1 \sim 4^{2}}$ or ${1 \sim 15^{2}}$, the effective slit force ranges from $1 \sim 4^{3}$ or $1 \sim 15^{3}$, and $t_{I}$ were set to 20 or 2, respectively. Additionally, we set $W_{L1}$ and $W_{R1}$ to take discrete values of ${1 \sim 4^{3}}$, with the slit force configuration fixed at $4^3$, and varied $t_{I}$ as 3, 6, 9, and 12. The corresponding simulation results are shown in Fig. 6(c$\sim$f), from which it is evident that  the emergence of interference fringes is sensitive to the choice of $t_{I}$ .

\begin{figure}
\centering
\includegraphics[width=8.5cm]{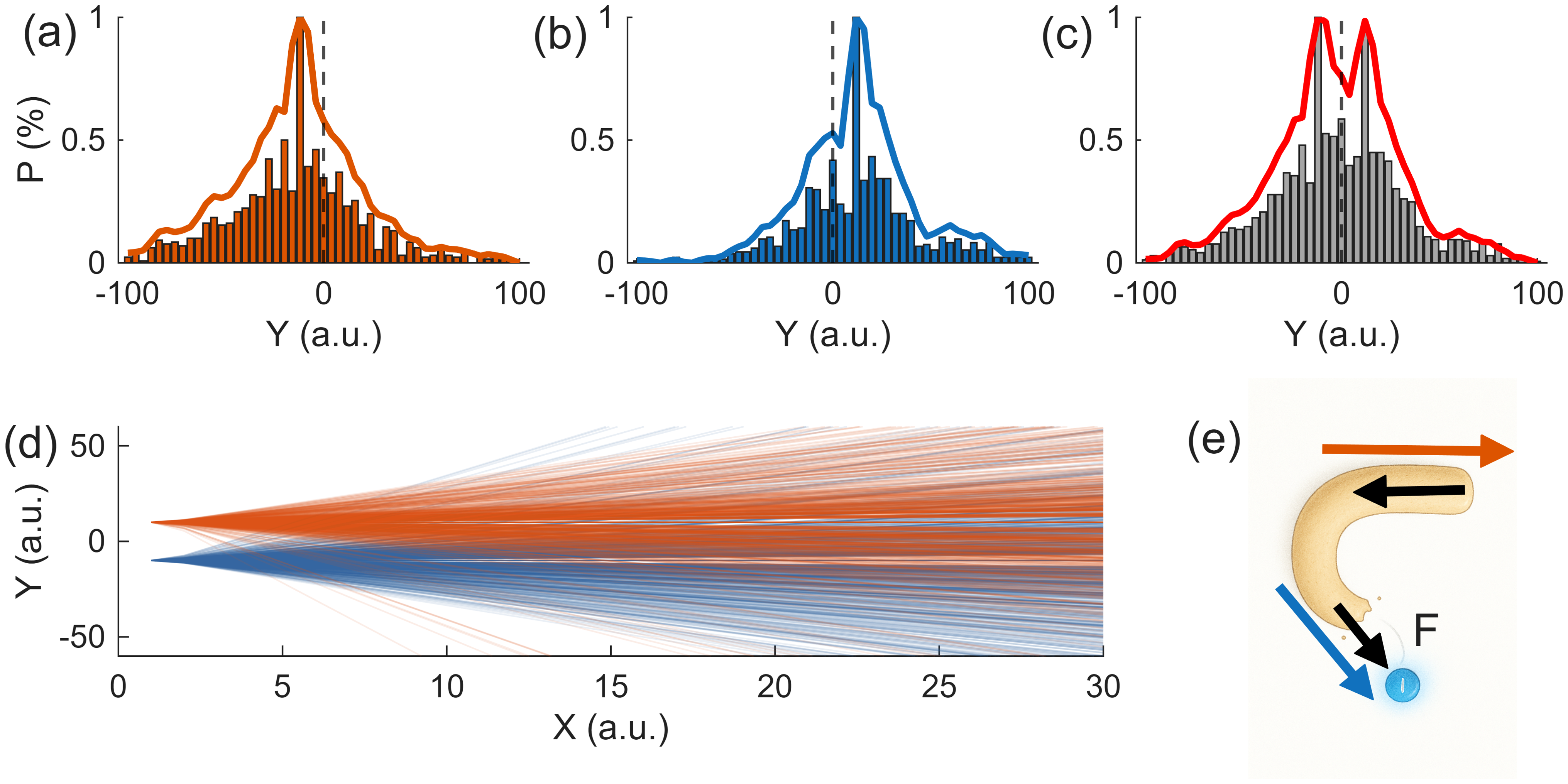}\caption{(a) The probability distribution from the left slit alone in the case of interference disruption, while (b) is from the right slit. (c) The sum of (a, b) without diffraction.
(d) The trajectories from (a) and (b) are represented by different color. 
(e) With interference disruption, the right branch of particle (indicated by the blue arrow) through the right slit reaches the detector marked by the blue point. The dangling left branch (red arrow) consequently retracts itself as if being pulled by a force $F$ (black arrow) toward the detector. }
\end{figure}

We also hypothesize a possible mechanism for destructive interference. 
First, we start from the two uncertainty relations
\begin{align}
  \Delta X\,\Delta P &\ge \frac{\hbar}{2},\\
  \Delta E\,\Delta T &\ge \frac{\hbar}{2}.
\end{align}
Define the normalized quantities
\begin{align}
  a &\equiv \frac{\Delta X\,\Delta P}{\hbar/2}\ge 1,\\
  b &\equiv \frac{\Delta E\,\Delta T}{\hbar/2}\ge 1.
\end{align}
The ratio of the two uncertainties can then be written as
\begin{align}
  \frac{\Delta X\,\Delta P}{\Delta E\,\Delta T}=\frac{a}{b}
  =\frac{\left(\frac{\Delta P}{\Delta T}\cdot \Delta X\right)}{\Delta E}
  =\frac{\Delta W}{\Delta E}.
\end{align}
which, without additional information about $a$ and $b$, may be greater than, equal to, or less than~1. 

Define $\Delta W$ as the work performed by the slit on the particle, and $\Delta E$ the corresponding energy change. The ratio $\Delta W / \Delta E$ therefore reflects the degree to which the slit’s action is effectively transferred to the particle’s energy. When $\Delta W / \Delta E = 1$, the work imparted by the slit exactly matches the particle’s energy change, indicating ideal conditions for quantum interference, where the  momentum change of the particle is properly aligned with its spatial uncertainty. Deviations from unity, i.e., $\Delta W / \Delta E > 1$ or $< 1$, correspond to additional perturbations: either excess or insufficient work applied to the particle. Such deviations disturb the particle’s motion, potentially degrading or even eliminating the interference pattern. Based on the above arguments, we present a picture in which the action of an external force 
disrupts the recombination of branches and renders the definition of  MP meaningless, while the additional perturbations increase the number of configurations of the applied force.


To simulate the destruction of interference, i.e., the reversion to two single-slit diffraction, we first increase the number of configuration to $15^3$ as a result of additional perturbations. 
If this number of configurations from slits is fixed, these additional configurations represent extra force variables due to external perturbations, which can raise or deduce the kinetic energy and prevent successful convergence at MP. Therefore, we use the positions of the two slits as the starting points instead.


We use the acceleration calculated from the forces  on the two channels divided by their respective stretched masses as a basis. A larger acceleration indicates a slower departure from the slit, causing that channel to reach the screen later than the other. 
Compared with the initial choice of 15, $t_{\rm I} = 2$ means a reduction in the time required for overlap at MP and interaction with the slits since we only focus on through which slit  the branch of dough arrives at the detector first.

Figures~7(a$\sim$c) show the histograms for 2000 instances passing through the left slit, right slit, and their sum, respectively. The trajectories are in Fig.~7(d), and the physical picture in Fig.~7(e). 
As shown in Fig. 2(g), the particle distribution deforms itself to $\subset$, donut, $\supset$, and back to spherical in order to pass through the double slits. If the 
evolution is cut short by the additional perturbations,  one branch of the $\subset$ may reach the detector before the donut configuration, i.e., no MP is formed. Subsequently, the other branch will be pulled back and toward the first one, as depicted in Fig. 7(e), which behavior is reminiscent of a dough. 











\section{Discussion}
Figures 1 and 2 illustrate our use of the DSM to establish multiple possible channels for connecting the wave function and the input–output relationships of the double-slit interference. Figure 1 presents the conceptual framework, while Fig. 2(f) shows the results of the encoding model. 
These results reveal that at three different time points the outputs on which the backbone networks rely represent the different optimal paths for wave function information transfer. 
Figure 2(g) further employs the “dough-like” model to summarize the possible transmission pathways of particles through the double slits. These findings provide a possible physical picture corresponding to (Q1) in Sec. III.

The t-SNE and centrality analyses in Fig. 3 further support the existence of a potential source after the slits, which contributes to the particle’s interference, consistent with the NL2 result shown in Fig. 2(g). This analysis provides a crucial insight for constructing a physical picture: when the “dough-like” structure gets stretched while passing through the slits, where does it recombine? Our picture suggests that the dough encompasses both slits, 
oozing  out of the slits two branches whose length can be different, and then these branches merge  allowing the mass they transport to recombine. The merge position can be defined by tracing their center-of-mass. Monte Carlo simulations account for the randomness arising from the vibrational fluctuations of the molecules or particles composing the slits, imparting stochastic deviations to the particle trajectory. 
Ultimately, Fig. 5 shows that our simulations are capable of reproducing the correct equally spaced interference fringes and the intensity distribution of diffraction. The outcomes under different parameter settings are discussed in Fig. 6.

In Fig. 7, we modified the Monte Carlo by removing the MP at NL2. The extensions and contractions of the two dough-like structures are coordinated - when one end reaches the detector or screen, the other end will react and get dragged toward the former. The density flow through the two slits is also discussed. In response to (Q2), this approach extends from the imagery derived in (Q1) and successfully provides a mechanism that produces both interference and non-interference phenomena under the stochastic Monte Carlo simulation. From a topological perspective, the incident particle can be regarded as exhibiting a topological state of $B^{2}(1)$. When stretched, it retains its topology until a loop $T_{1}$ is formed. However, the way this loop reconnects may cause the post-NL1 state $B^{2}(2)$ to become non-homeomorphic to $B^{2}(1)$, leading to different transverse direction. Conversely, if $B^{2}(1)$ and $B^{2}(2)$ are homeomorphic, no interference fringes will appear.

\begin{figure}
\centering
\includegraphics[width=8.5cm]{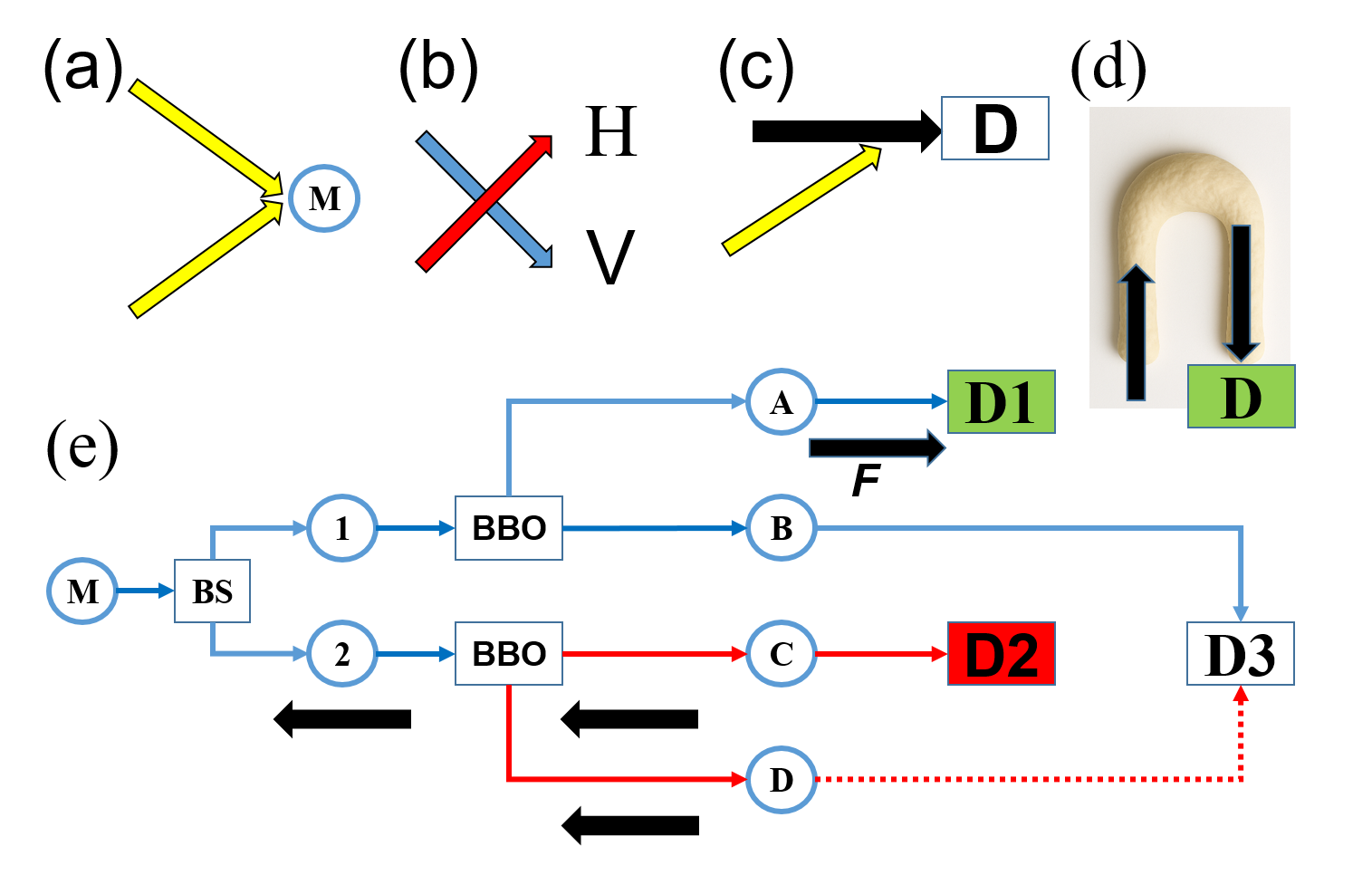}\caption{(a) Schematic plot for the recombination of branches that emerge from the double slits, with M representing the total mass. (b) The branches may fail to merge in special instances. (c) The black branch reaches the detector D faster than the yellow one, determining the measured position. Although the yellow branch eventually merges into the black branch, as in (a), it cannot affect the trajectory changes. (d) Similar scenario as (c) without the merger - the second branch simply gets dragged along by the first one and got captured later by the detector. 
(e) Flow chart for the which-way experiment. Since BBO is a nonlinear crystal generating entangled photon pairs and BS is a beam splitter, a head is formed at that location when D1 receives the stretched quantity, D2 does not receive any signal. The absorbed action associated with this branch influences branches C and D along path 2 to arrive later than branch B along path 1. As a result, we always conclude that the photon travels along path 1.}
\end{figure}  

The Monte Carlo framework was based on the conclusions drawn from the DSM. NL1 serves as the coupling point that links the two extension--contraction components passing through the respective slits, maintaining a continuous connection without invoking any concept of duplication. Both sides, however, are permitted to transmit fractional masses. We hypothesize that this transmission mechanism originates from the positional configurations determined by the intrinsic uncertainty in the atomic positions of the slit material. 

Although these configurations form a fixed ensemble of possible events, stochastic selection within each time interval introduces uncertainty, thereby giving rise to an effective quantum potential. Under such configurations, the slits or gratings induce differential magnitudes of extension on the two branches. 
Before reaching the center-of-mass position defined by NL2, these branches accumulate distinct force contributions. Since NL2 represents the center-of-mass, it inherits all the forces. When NL1 finally ruptures and all fractional masses recombine at NL2 into a topology denoted as 
\(
B^{2}(2)
\),
the particle trajectory is consequently modified. 

From a physics standpoint, the emergence of such specialized topological channels - constructed through global geometric phases 
or system-level winding numbers is not uncommon \cite{d2020two,tarnowski2019measuring,schroter2020observation}.
Based on the dough model, the uncertainty principle is not difficult to understand. In terms of momentum and spatial uncertainties, detecting the position means that a certain segment of the stretched space has been intercepted—for example, when photons enter a CCD and undergo photoelectron conversion \cite{pratt1973atomic}, this point becomes the end of the dough, yielding a definite position. However, the uncertainty in momentum increases, because the true momenta of the various moving branches cannot be obtained. If we instead consider the slit-and-track setup described in Richard Feynman’s big red book, where particle momentum is measured after collisions on the sliding track, the momentum of the track reflects the collective result of the two slit branches. Yet, sliding the slit affects the original arrival positions at NL2, and may even prevent NL2 from effectively merging the two stretched segments. Consequently, the subsequent trajectories cannot propagate to the screen according to the set of Monte Carlo events prescribed by the slits, and what we observe is blurred noise rather than interference fringes.


The effect of the slit on the quantum entity proposed in our model is similar to the quantum potential in Bohmian mechanics, i.e., there is no feedback effect on the slits. However, a key distinction lies in that our model allows the particle to respond to the slits only in their close vicinity. 
The configuration of atoms or molecules on the slits and the resulting force on the particle governs the relative volume or mass of branches sticking out of the double slits and their consequent trajectory. In correspondence with the view on evolution by John von Neumann\cite{von2018mathematical}, first-type evolution corresponds to the behavior of NL2 under randomly assigned set of Monte Carlo events, while second-type evolution to the process in which the “dough” preferentially reaches certain regions and undergoes collective contraction.

The trajectory of the dough in the physical picture remains globally continuous and never breaks, while allowing for backflow, twisting, and bending. These characteristics have been discussed in near-field optics and previous studies \cite{magana2016exotic,sawant2013non}. In our simulations, the MP generation points are predominantly located between the two slits, closely resembling the near-field diffraction observed in numerical simulations \cite{teng2013near}. Furthermore, experiments in which a double slit evolves into a triple slit due to plasmon-induced bending of light also demonstrate that the slits serve as the primary modulators of information transfer \cite{magana2016exotic,schouten2005plasmon}, which inspired our DSM framework and Monte Carlo simulations.

Our theory differs from the standard quantum mechanics based on statistical ensembles; it is instead a description grounded in the behavior of single quanta. Here, wave-like behavior represents trajectories sampled from the set of Monte Carlo events, whereas particle-like behavior corresponds to the detection location registered by the detector. Where our approach aligns with the interpretation of quantum mechanics is that interactions between the instrument and the system indeed modify subsequent behavior. 
According to the simulation for Fig. 7, a cut short and stop at the second stage occurs without the formation of NL2 in Fig. 2(g), i.e., once a branch reached the detector, the other branch would react and got dragged along as if the particle is elastic. 


In traditional hidden variable theories \cite{paul1980einstein,walborn2011revealing}, a particle’s properties are fully determined by parameters confined to its immediate vicinity, which cannot be influenced by events occurring at distant locations. However, experimental results, such as violations of Bell’s inequalities \cite{aspect1982experimental}, indicate that strict locality does not fully describe nature. In contrast, nonlocal hidden variables posit that the state of a particle is governed by correlations within a global system, such that its internal structure can be constrained simultaneously by the overall field, even when spatially separated. 

The similitude to dough treats a quantum particle not as a point-like entity but as a continuous, deformable, and extended structure. Both interference and collapse phenomena can be interpreted as natural processes of deformation and reconfiguration of this extended entity within space and time. Upon detection, the system does not rely on instantaneous signal transmission; instead, the entire dough-like structure readjusts globally to reach a stable configuration. Consequently, the particle’s behavior is determined not by independent local parameters but by the internal correlations of the whole field across space and time, aligning with the principles of nonlocal hidden variable theories \cite{healey1997nonlocality,peres1997quantum}.


It should be emphasized that the dough model is hypothetical. While Monte Carlo simulations have successfully reproduced both interference and the transition to non-interference in a double-slit setup, experimental verification remains outstanding. Therefore, the conclusions presented here should be regarded as a theoretical proposal, offering a new perspective on quantum phenomena from the viewpoint of nonlocal hidden variables and realism. With future experiments designed from this perspective, the dough model may potentially serve as a framework for a realist interpretation of quantum theory.

Combining the dough analogy and Monte Carlo simulations for the double-slit experiment, we deduce that: (1) based on the indivisibility guaranteed by the dough image, the two branches of the photon or electron that stick out from the double slits are stretched separately, which more usual than not results in the variance of their volume or mass. 
(2) once the two branches recombine after the slits, their other ends before the slits are allowed to split without dividing the particle in half. 
(3) After the reunification, the information of forces received by individual branch locally got integrated and thus became nonlocal, as illustrated in Fig.~8(a). This is analogous to the Aharonov–Bohm effect, where the 
sum of the time-evolution operator from the two slits results in the difference of their phases which equals a loop integral of the vector potentials or equivalently, by use of the Stokes' theorem, the magnetic flux enclosed by these two paths.



Figure 8(b) illustrates another scenario, depicting how our theoretical framework accounts for the disruption of interference fringes in the double-slit experiment due to perturbations: when path information, such as polarization, is assigned, the structure may lose the NL2 connection, similar to a quantum eraser experiment, resulting in the disappearance of interference fringes. Attempting to label the path information can introduce differences in the intrinsic stretching properties of the two branches, preventing them from merging into a global structure that preserves the histories of both paths, as demonstrated in our Monte Carlo simulations in Fig.~7: the branch that reaches the detector first becomes the "head," while the other becomes the "tail," as illustrated in Fig.~8(c) and (d). 


Should the dough analogy be in line with the fundamental logic of quantum mechanics, it is likely to also provide a framework for understanding the entanglement and tunneling phenomena. Entanglement \cite{horodecki2009quantum} characterizes the inseparability of a composite quantum state, meaning that the total system cannot be decomposed into independent subsystems. This structural feature does not imply any physical signal being transmitted instantaneously. Nonlocality, in the strict sense, refers to the violation of Bell inequalities, which represents a stronger condition than entanglement and cannot be explained by any local hidden-variable theory. 

The two branches that emerge from the double slits are not independent in our dough model, instead are components of a single and unified quantum state. Their correlations arise from the global structure of this joint state rather than from classical information transfer between spatially separated parts. This perspective highlights the intrinsic, nonclassical organization of an entangled system without invoking any faster-than-light influence.



Based on the above perspective, we consider that a pair of entangled photons generated by Beta Barium Borate (BBO) \cite{amselem2009experimental} can still be regarded as a single “dough,” though they are somewhat special in that they are adhered together. Entanglement, in the context of scientific realism, corresponds to the possibility of performing measurement operations on both photons, ensuring that at least two measurements can be carried out. The entanglement is constituted by the internal connections and structure of the dough’s stretching, encompassing all the information of the Hilbert space, as well as the configuration space in which the information resides.

Generalizing the dough analogy to the photon pair entangled via BBO, the two photons resemble a whole loaf of bread with two types of filling: if one randomly obtains one type  at one end (e.g., peanut), the measurement at the other end can only yield the remaining filling (say, red bean). When detector A determines the red bean filling, detector B can never obtain that same filling. Therefore, the “unmeasurable part” can be regarded as substituting for the measurement information we focus on; that is, it is not that the peanut filling instantaneously reaches detector B, but rather that the red bean filling gradually moves away from detector A, until the two portions of dough are progressively pulled and ultimately separated under the influence of two detectors. If polarization information is obtained at both ends, this does not imply that the filling has been fully transmitted; rather, it has not yet been entirely absorbed by the detectors before separation. The entire process still obeys the event horizon determined by the speed of light, thus remaining consistent with classical causality. Within this framework, measurements of entanglement involve no classical information transfer and instead manifest as nonlocal correlations of the global state.

This provides a clear understanding of the results of the which-way experiment \cite{durr1998origin,walborn2002double,herzog1995complementarity}, as illustrated in Fig. 8(e). The two paths after the beam splitter, paths 1 and 2, form a single coherent whole. When path A is detected by D1, a detector-induced attraction is exerted, causing the branch along path 2 in the BS to act as the tail, while the B branch along path 1 reaches detector D3. If paths A and C are removed, the system returns to the outcome shown in Fig. 8(a). The same reasoning applies to the quantum eraser experiment in a Mach-Zehnder interferometer. In the scenario depicted in Fig. 8(b), the elimination of path polarization information can similarly restore the outcome to that of Fig. 8(a).


If we interpret tunneling \cite{merzbacher2002early} from the perspective of the configuration used in the double-slit diffraction, the potential landscape should also vary spatially according to the material structure. This makes it possible for the “dough” to branch out, similar to the percolation theory\cite{shante1971introduction,essam1980percolation}, where multiple branches may merge or further split. The branch that eventually reaches the exit may therefore no longer align with its point of entry, producing a lateral displacement. This lateral displacement could also manifest itself in reflection as the Goos–Hänchen shift\cite{snyder1976goos}.

\begin{figure}
\centering
\includegraphics[width=9cm]{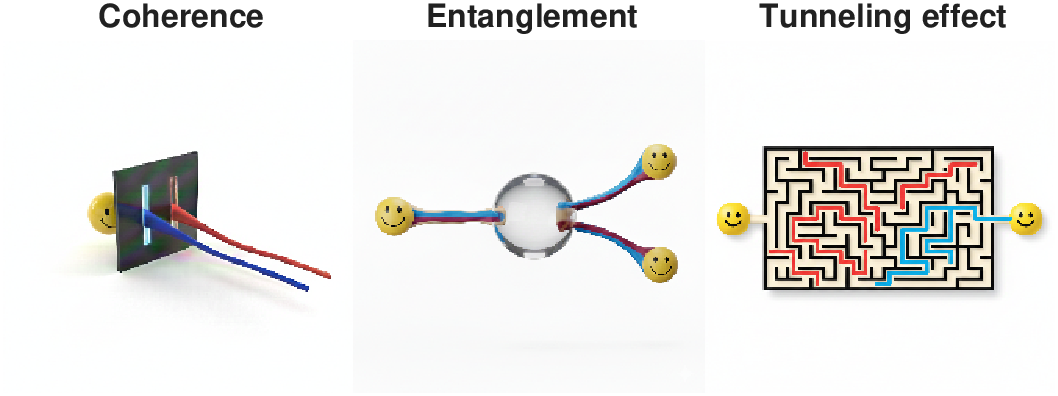}\caption{The three peculiar quantum phenomena, based on the dough model, allow a realist description through stretching, corresponding to interference and diffraction. In the entanglement picture, the Hilbert space basis is contained like a filling: when one party randomly obtains a certain filling, the other party inevitably receives the remaining filling. Tunneling is analogous to the branching of many stretched dough strands, which, upon reaching the exit, retain the ability to stretch and influence the other branches.}
\end{figure}

The above Monte Carlo simulation represents a simplified conceptual model. In the future, it will be necessary to compare with the arrival-time distributions \cite{ayatollah2023can,kazemi2023detection} to further refine and validate the model. Moreover, the material parameters underlying the nonlocal elastomer in the dough model require further verification. In this study, we also use the illustrations in Fig. 9 to  interpret three  quantum characteristics based on the theoretical imagery. Additionally, whether the coherence length imposes a condition for the two branches of dough to form an effective NL2 is another direction worth investigating\cite{koushki2023study, redding2011spatial}. Based on our theory, both quantum ghost imaging and classical ghost imaging \cite{erkmen2010ghost,padgett2017introduction} arise from the same underlying mechanism under different conditions, and our framework may potentially provide insights for interpreting these phenomena.

\section{Conclusion}
In this study, we have introduced the dough model as a conceptual and computational framework to describe the behavior of a single quantum in the double-slit experiment, in combination with stochastic Monte Carlo simulations. Our results indicate that the extended and deformable nature of the quantum entity enables simultaneous traversal through both slits, recombination of masses, and the emergence of both interference and non-interference patterns. 

As a new interpretation grounded in scientific realism, we aim to complement the projection postulate by explicitly describing the intermediate processes occurring from measurement to collapse, while also suggesting new avenues for physically motivated experimental designs. In particular, our reproduction of double-slit interference arises from allowing the mass of the branches traversing the slits to be distributed proportionally, rather than invoking wavefunction splitting in the Copenhagen interpretation or forcing the quantum into a binary, either–or outcome at the slits.

The dough model also try to offer an unifying alternative perspective on quantum correlations, entanglement, and tunneling. The global reconfiguration of the dough upon detection suggests a mechanism for nonlocal correlations that does not require classical information transfer. By treating particles as extended entities with topologically constrained trajectories, the dough model emphasizes the importance of global geometric and topological properties in quantum processes. While it remains unclear whether the dough model is directly related to dark matter or ER=EPR phenomena \cite{jafferis2022stringy,susskind2016er,dai2020testing}, the extended structure and nonlocal behavior may provide a offbeat perspective for exploring connections between quantum phenomena, spacetime geometry, and currently unknown physical components of the universe.


\section{ACKNOWLEDGMENTS}
We are grateful to the financial support from the National Science and Technology Council in Taiwan under Grants No. 113-2112-M007-008 and 114-2112-M007-004.



\bibliography{apssamp}

@article{felippa2001historical,
  title={A historical outline of matrix structural analysis: a play in three acts},
  author={Felippa, Carlos A},
  journal={Computers \& Structures},
  volume={79},
  number={14},
  pages={1313--1324},
  year={2001},
  publisher={Elsevier}
}

@article{levada2018review,
  title={Review of the Schr{\"o}dinger Wave Equation},
  author={Levada, Celso Luis and Maceti, Huemerson and Lautenschleguer, Ivan Jos{\'e}},
  journal={IOSR Journal of Applied Chemistry},
  volume={11},
  number={4},
  pages={01--07},
  year={2018}
}

@incollection{bouwmeester2000physics,
  title={The physics of quantum information: basic concepts},
  author={Bouwmeester, Dirk and Zeilinger, Anton},
  booktitle={The physics of quantum information: quantum cryptography, quantum teleportation, quantum computation},
  pages={1--14},
  year={2000},
  publisher={Springer}
}

@article{durr1998origin,
  title={Origin of quantum-mechanical complementarity probed by a ‘which-way’experiment in an atom interferometer},
  author={D{\"u}rr, S and Nonn, T and Rempe, G},
  journal={Nature},
  volume={395},
  number={6697},
  pages={33--37},
  year={1998},
  publisher={Nature Publishing Group UK London}
}

@article{walborn2002double,
  title={Double-slit quantum eraser},
  author={Walborn, SP and Cunha, MO Terra and P{\'a}dua, S and Monken, CH},
  journal={Physical Review A},
  volume={65},
  number={3},
  pages={033818},
  year={2002},
  publisher={APS}
}

@book{gibbins1987particles,
  title     = {Particles and Paradoxes: The Limits of Quantum Logic},
  author    = {Gibbins, Peter},
  publisher = {Cambridge University Press},
  address   = {Cambridge},
  year      = {1987}
}

@article{gao2011meaning,
  title={Meaning of the wave function},
  author={Gao, Shan},
  journal={International Journal of Quantum Chemistry},
  volume={111},
  number={15},
  pages={4124--4138},
  year={2011},
  publisher={Wiley Online Library}
}

@article{goldstein2013reality,
  title={Reality and the role of the wave function in quantum theory},
  author={Goldstein, Sheldon and Zangh{\`\i}, Nino},
  journal={The wave function: Essays on the metaphysics of quantum mechanics},
  pages={91--109},
  year={2013}
}

@article{ballentine1970statistical,
  title={The statistical interpretation of quantum mechanics},
  author={Ballentine, Leslie E},
  journal={Reviews of modern physics},
  volume={42},
  number={4},
  pages={358},
  year={1970},
  publisher={APS}
}

@article{schlegel1970statistical,
  title={Statistical explanation in physics: The Copenhagen interpretation},
  author={Schlegel, Richard},
  journal={Synthese},
  volume={21},
  number={1},
  pages={65--82},
  year={1970},
  publisher={Springer}
}

@article{kaiser2014history,
  title={History: Shut up and calculate!},
  author={Kaiser, David},
  journal={Nature},
  volume={505},
  number={7482},
  pages={153--155},
  year={2014},
  publisher={Nature Publishing Group UK London}
}

@article{adlam2025kind,
  title={What Kind of Relationality does Quantum Mechanics Exhibit?},
  author={Adlam, Emily},
  journal={arXiv preprint arXiv:2502.06991},
  year={2025}
}

@article{bell1990against,
  title={Against ‘measurement’},
  author={Bell, John},
  journal={Physics world},
  volume={3},
  number={8},
  pages={33},
  year={1990},
  publisher={IOP Publishing}
}

@article{einstein1935can,
  title={Can quantum-mechanical description of physical reality be considered complete?},
  author={Einstein, Albert and Podolsky, Boris and Rosen, Nathan},
  journal={Physical review},
  volume={47},
  number={10},
  pages={777},
  year={1935},
  publisher={APS}
}

@book{albert2009quantum,
  title={Quantum mechanics and experience},
  author={Albert, David Z},
  year={2009},
  publisher={Harvard University Press}
}

@article{natarajan2008einstein,
  title={What Einstein meant when he said “God does not play dice…”},
  author={Natarajan, Vasant},
  journal={Resonance},
  volume={13},
  number={7},
  pages={655--661},
  year={2008},
  publisher={Springer}
}

@inproceedings{kwiat1994experimental,
  title={Experimental realization of “Interaction-Free” measurements},
  author={Kwiat, Paul and Weinfurter, Harald and Herzog, Thomas and Zeilinger, Anton and Kasevich, Mark},
  booktitle={Fundamental Problems in Quantum Theory: A Conference Held in Honor ofProfessor John A. Wheeler},
  volume={755},
  pages={129},
  year={1994}
}

@incollection{robens2018atomic,
  title={Atomic “bomb testing”: the Elitzur--Vaidman experiment violates the Leggett--Garg inequality},
  author={Robens, Carsten and Alt, Wolfgang and Emary, Clive and Meschede, Dieter and Alberti, Andrea},
  booktitle={Exploring the World with the Laser: Dedicated to Theodor H{\"a}nsch on his 75th birthday},
  pages={141--157},
  year={2018},
  publisher={Springer}
}

@article{wiseman1996quantum,
  title={Quantum trajectories and quantum measurement theory},
  author={Wiseman, Howard M},
  journal={Quantum and Semiclassical Optics: Journal of the European Optical Society Part B},
  volume={8},
  number={1},
  pages={205},
  year={1996},
  publisher={IOP Publishing}
}

@article{svensson2013pedagogical,
  title={Pedagogical review of quantum measurement theory with an emphasis on weak measurements},
  author={Svensson, Bengt EY},
  journal={Quanta},
  volume={2},
  number={1},
  pages={18--49},
  year={2013},
  publisher={Quanta}
}

@article{foo2022relativistic,
  title={Relativistic Bohmian trajectories of photons via weak measurements},
  author={Foo, Joshua and Asmodelle, Estelle and Lund, Austin P and Ralph, Timothy C},
  journal={Nature communications},
  volume={13},
  number={1},
  pages={4002},
  year={2022},
  publisher={Nature Publishing Group UK London}
}

@article{matzkin2015weak,
  title={Weak measurements of trajectories in quantum systems: classical, Bohmian and sum over paths},
  author={Matzkin, Alex},
  journal={Journal of Physics A: Mathematical and Theoretical},
  volume={48},
  number={30},
  pages={305301},
  year={2015},
  publisher={IOP Publishing}
}

@article{mahler2016experimental,
  title={Experimental nonlocal and surreal Bohmian trajectories},
  author={Mahler, Dylan H and Rozema, Lee and Fisher, Kent and Vermeyden, Lydia and Resch, Kevin J and Wiseman, Howard M and Steinberg, Aephraim},
  journal={Science advances},
  volume={2},
  number={2},
  pages={e1501466},
  year={2016},
  publisher={American Association for the Advancement of Science}
}

@article{mena2021solving,
  title={Solving the 2D Schr{\"o}dinger equation using the Crank-Nicolson method},
  author={Mena, Arturo},
  journal={Quantum Things},
  volume={26},
  year={2021}
}

@misc{artmenlope2021double-slit,
  author = {Arturo Mena López},
  title = {double-slit-2d-schrodinger: Python scripts for simulating 2D Gaussian wave packet through a double slit},
  year = {2021},
  publisher = {GitHub},
  journal = {GitHub repository},
  howpublished = {\url{https://github.com/artmenlope/double-slit-2d-schrodinger}},
  note = {MIT License}
}

@article{werner2001bell,
  title={Bell inequalities and entanglement},
  author={Werner, Reinhard F and Wolf, Michael M},
  journal={arXiv preprint quant-ph/0107093},
  year={2001}
}

@article{anderson1963probable,
  title={Probable observation of the Josephson superconducting tunneling effect},
  author={Anderson, Philip W and Rowell, John M},
  journal={Physical Review Letters},
  volume={10},
  number={6},
  pages={230},
  year={1963},
  publisher={APS}
}

@article{tonomura1986evidence,
  title={Evidence for Aharonov-Bohm effect with magnetic field completely shielded from electron wave},
  author={Tonomura, Akira and Osakabe, Nobuyuki and Matsuda, Tsuyoshi and Kawasaki, Takeshi and Endo, Junji and Yano, Shinichiro and Yamada, Hiroji},
  journal={Physical review letters},
  volume={56},
  number={8},
  pages={792},
  year={1986},
  publisher={APS}
}

@article{osakabe1986experimental,
  title={Experimental confirmation of Aharonov-Bohm effect using a toroidal magnetic field confined by a superconductor},
  author={Osakabe, Nobuyuki and Matsuda, Tsuyoshi and Kawasaki, Takeshi and Endo, Junji and Tonomura, Akira and Yano, Shinichiro and Yamada, Hiroji},
  journal={Physical Review A},
  volume={34},
  number={2},
  pages={815},
  year={1986},
  publisher={APS}
}

@article{arndt1999wave,
  title={Wave--particle duality of C60 molecules},
  author={Arndt, Markus and Nairz, Olaf and Vos-Andreae, Julian and Keller, Claudia and Van der Zouw, Gerbrand and Zeilinger, Anton},
  journal={nature},
  volume={401},
  number={6754},
  pages={680--682},
  year={1999},
  publisher={Nature Publishing Group UK London}
}

@article{magana2016exotic,
  title={Exotic looped trajectories of photons in three-slit interference},
  author={Magana-Loaiza, Omar S and De Leon, Israel and Mirhosseini, Mohammad and Fickler, Robert and Safari, Akbar and Mick, Uwe and McIntyre, Brian and Banzer, Peter and Rodenburg, Brandon and Leuchs, Gerd and others},
  journal={Nature Communications},
  volume={7},
  number={1},
  pages={13987},
  year={2016},
  publisher={Nature Publishing Group UK London}
}

@article{sorkin1994quantum,
  title={Quantum mechanics as quantum measure theory},
  author={Sorkin, Rafael D},
  journal={Modern Physics Letters A},
  volume={9},
  number={33},
  pages={3119--3127},
  year={1994},
  publisher={World Scientific}
}

@article{sinha2010ruling,
  title={Ruling out multi-order interference in quantum mechanics},
  author={Sinha, Urbasi and Couteau, Christophe and Jennewein, Thomas and Laflamme, Raymond and Weihs, Gregor},
  journal={Science},
  volume={329},
  number={5990},
  pages={418--421},
  year={2010},
  publisher={American Association for the Advancement of Science}
}

@article{villas2025bright,
  title={Bright and dark states of light: The quantum origin of classical interference},
  author={Villas-Boas, Celso J and M{\'a}ximo, Carlos E and Paulino, Paulo J and Bachelard, Romain P and Rempe, Gerhard},
  journal={Physical Review Letters},
  volume={134},
  number={13},
  pages={133603},
  year={2025},
  publisher={APS}
}

@article{fedoseev2025coherent,
  title={Coherent and incoherent light scattering by single-atom wave packets},
  author={Fedoseev, Vitaly and Lin, Hanzhen and Lu, Yu-Kun and Lee, Yoo Kyung and Lyu, Jiahao and Ketterle, Wolfgang},
  journal={Physical Review Letters},
  volume={135},
  number={4},
  pages={043601},
  year={2025},
  publisher={APS}
}

@incollection{planck1978gesetz,
  title={{\"U}ber das gesetz der energieverteilung im normalspektrum},
  author={Planck, Max},
  booktitle={Von Kirchhoff bis Planck: Theorie der W{\"a}rmestrahlung in historisch-kritischer Darstellung},
  pages={178--191},
  year={1978},
  publisher={Springer}
}

@article{gibney2025physicists,
  title={Physicists disagree wildly on what quantum mechanics says about reality, Nature survey shows},
  author={Gibney, Elizabeth},
  journal={Nature},
  volume={643},
  number={8074},
  pages={1175--1179},
  year={2025}
}

@article{sood2023quantum,
  title={Quantum computing review: A decade of research},
  author={Sood, Sandeep Kumar and others},
  journal={IEEE Transactions on Engineering Management},
  volume={71},
  pages={6662--6676},
  year={2023},
  publisher={IEEE}
}

@article{gyongyosi2019survey,
  title={A survey on quantum computing technology},
  author={Gyongyosi, Laszlo and Imre, Sandor},
  journal={Computer Science Review},
  volume={31},
  pages={51--71},
  year={2019},
  publisher={Elsevier}
}

@article{wei2022towards,
  title={Towards real-world quantum networks: a review},
  author={Wei, Shi-Hai and Jing, Bo and Zhang, Xue-Ying and Liao, Jin-Yu and Yuan, Chen-Zhi and Fan, Bo-Yu and Lyu, Chen and Zhou, Dian-Li and Wang, You and Deng, Guang-Wei and others},
  journal={Laser \& Photonics Reviews},
  volume={16},
  number={3},
  pages={2100219},
  year={2022},
  publisher={Wiley Online Library}
}

@article{bohm1961hidden,
  title={Hidden variables in the quantum theory},
  author={Bohm, David},
  journal={Quantum theory},
  volume={3},
  pages={345--387},
  year={1961}
}

@article{albert1988interpreting,
  title={Interpreting the many worlds interpretation},
  author={Albert, David and Loewer, Barry},
  journal={Synthese},
  pages={195--213},
  year={1988},
  publisher={JSTOR}
}

@article{kolenderski2014time,
  title={Time-resolved double-slit interference pattern measurement with entangled photons},
  author={Kolenderski, Piotr and Scarcella, Carmelo and Johnsen, Kelsey D and Hamel, Deny R and Holloway, Catherine and Shalm, Lynden K and Tisa, Simone and Tosi, Alberto and Resch, Kevin J and Jennewein, Thomas},
  journal={Scientific reports},
  volume={4},
  number={1},
  pages={4685},
  year={2014},
  publisher={Nature Publishing Group UK London}
}

@article{holland1993broglie,
  title={The de Broglie-Bohm theory of motion and quantum field theory},
  author={Holland, Peter R},
  journal={Physics reports},
  volume={224},
  number={3},
  pages={95--150},
  year={1993},
  publisher={Elsevier}
}

@article{plaga1997possibility,
  title={On a possibility to find experimental evidence for the many-worlds interpretation of quantum mechanics},
  author={Plaga, Rainer},
  journal={Foundations of Physics},
  volume={27},
  number={4},
  pages={559--577},
  year={1997},
  publisher={Springer}
}

@article{timpson2008quantum,
  title={Quantum Bayesianism: a study},
  author={Timpson, Christopher Gordon},
  journal={Studies in History and Philosophy of Science Part B: Studies in History and Philosophy of Modern Physics},
  volume={39},
  number={3},
  pages={579--609},
  year={2008},
  publisher={Elsevier}
}

@article{shapiro1989aharonov,
  title={The Aharonov-Bohm effect in double-and single-slit diffraction},
  author={Shapiro, D and Henneberger, WC},
  journal={Journal of Physics A: Mathematical and General},
  volume={22},
  number={17},
  pages={3605},
  year={1989},
  publisher={IOP Publishing}
}

@article{li2021survey,
  title={A survey of convolutional neural networks: analysis, applications, and prospects},
  author={Li, Zewen and Liu, Fan and Yang, Wenjie and Peng, Shouheng and Zhou, Jun},
  journal={IEEE transactions on neural networks and learning systems},
  volume={33},
  number={12},
  pages={6999--7019},
  year={2021},
  publisher={IEEE}
}

@article{serrano2009extracting,
  title   = {Extracting the multiscale backbone of complex weighted networks},
  author  = {Serrano, M. {\'A}ngeles and Bogun{\'a}, Mari{\'a}n and Vespignani, Alessandro},
  journal = {Proceedings of the National Academy of Sciences},
  volume  = {106},
  number  = {16},
  pages   = {6483--6488},
  year    = {2009},
  publisher = {National Academy of Sciences}
}

@article{simas2021distance,
  title={The distance backbone of complex networks},
  author={Simas, Tiago and Correia, Rion Brattig and Rocha, Luis M},
  journal={Journal of Complex Networks},
  volume={9},
  number={6},
  pages={cnab021},
  year={2021},
  publisher={Oxford University Press}
}

@article{eickenberg2017seeing,
  title={Seeing it all: Convolutional network layers map the function of the human visual system},
  author={Eickenberg, Michael and Gramfort, Alexandre and Varoquaux, Ga{\"e}l and Thirion, Bertrand},
  journal={NeuroImage},
  volume={152},
  pages={184--194},
  year={2017},
  publisher={Elsevier}
}

@article{guo2020first,
  title={The first principles for artificial intelligence},
  author={Guo, Ping},
  journal={Communications of the China Computer Federation},
  volume={16},
  number={10},
  pages={53--58},
  year={2020}
}

@article{aharonov1959significance,
  title={Significance of electromagnetic potentials in the quantum theory},
  author={Aharonov, Yakir and Bohm, David},
  journal={Physical review},
  volume={115},
  number={3},
  pages={485},
  year={1959},
  publisher={APS}
}

@article{lounis2005single,
  title={Single-photon sources},
  author={Lounis, Brahim and Orrit, Michel},
  journal={Reports on Progress in Physics},
  volume={68},
  number={5},
  pages={1129},
  year={2005},
  publisher={IOP Publishing}
}

@article{yuen2024exact,
  title={Exact quantum electrodynamics of radiative photonic environments},
  author={Yuen, Ben and Demetriadou, Angela},
  journal={Physical review letters},
  volume={133},
  number={20},
  pages={203604},
  year={2024},
  publisher={APS}
}

@book{dirac1981principles,
  title={The principles of quantum mechanics},
  author={Dirac, Paul Adrien Maurice},
  number={27},
  year={1981},
  publisher={Oxford university press}
}

@article{teller1983projection,
  title={The projection postulate as a fortuitous approximation},
  author={Teller, Paul},
  journal={Philosophy of Science},
  volume={50},
  number={3},
  pages={413--431},
  year={1983},
  publisher={Cambridge University Press}
}

@article{margenau1936quantum,
  title={Quantum-mechanical description},
  author={Margenau, Henry},
  journal={Physical Review},
  volume={49},
  number={3},
  pages={240--242},
  year={1936}
}

@book{bell2004speakable,
  title={Speakable and unspeakable in quantum mechanics: Collected papers on quantum philosophy},
  author={Bell, John Stewart},
  year={2004},
  publisher={Cambridge university press}
}

@book{cushing1994quantum,
  title={Quantum mechanics: historical contingency and the Copenhagen hegemony},
  author={Cushing, James T},
  year={1994},
  publisher={University of Chicago Press}
}

@article{landsman2006between,
  title={Between classical and quantum},
  author={Landsman, Nicolaas P},
  journal={Handbook of the Philosophy of Science},
  volume={2},
  pages={417--553},
  year={2006}
}

@book{baggott2004beyond,
  title={Beyond measure: Modern physics, philosophy, and the meaning of quantum theory},
  author={Baggott, James Edward},
  year={2004},
  publisher={Oxford University Press}
}

@book{berezin2012schrodinger,
  title={The Schr{\"o}dinger Equation},
  author={Berezin, Feliks Aleksandrovich and Shubin, Mikhail},
  volume={66},
  year={2012},
  publisher={Springer Science \& Business Media}
}

@article{schrodinger1935present,
  title={The present status of quantum mechanics},
  author={Schr{\"o}dinger, Erwin},
  journal={Die Naturwissenschaften},
  volume={23},
  number={48},
  pages={1--26},
  year={1935}
}

@book{von2018mathematical,
  title={Mathematical foundations of quantum mechanics: New edition},
  author={Von Neumann, John},
  year={2018},
  publisher={Princeton university press}
}

@book{von2013mathematische,
  title={Mathematische grundlagen der quantenmechanik},
  author={Von Neumann, John},
  volume={38},
  year={2013},
  publisher={Springer-Verlag}
}

@book{aerts2013quantum,
  title={Quantum Structures and the Nature of Reality: The Indigo Book ofEinstein Meets Magritte'},
  author={Aerts, Diederik and Pykacz, Jaroslaw},
  volume={7},
  year={2013},
  publisher={Springer Science \& Business Media}
}

@article{peres1999all,
  title={All the Bell inequalities},
  author={Peres, Asher},
  journal={Foundations of Physics},
  volume={29},
  number={4},
  pages={589--614},
  year={1999},
  publisher={Springer}
}

@article{blaylock2010epr,
  title={The EPR paradox, Bell’s inequality, and the question of locality},
  author={Blaylock, Guy},
  journal={American Journal of Physics},
  volume={78},
  number={1},
  pages={111--120},
  year={2010},
  publisher={AIP Publishing}
}

@article{busch2007heisenberg,
  title={Heisenberg's uncertainty principle},
  author={Busch, Paul and Heinonen, Teiko and Lahti, Pekka},
  journal={Physics reports},
  volume={452},
  number={6},
  pages={155--176},
  year={2007},
  publisher={Elsevier}
}

@article{lecun2015deep,
  title={Deep learning},
  author={LeCun, Yann and Bengio, Yoshua and Hinton, Geoffrey},
  journal={nature},
  volume={521},
  number={7553},
  pages={436--444},
  year={2015},
  publisher={Nature Publishing Group UK London}
}

@article{liu2021global,
  title={A global surrogate model technique based on principal component analysis and Kriging for uncertainty propagation of dynamic systems},
  author={Liu, Yushan and Li, Luyi and Zhao, Sihan and Song, Shufang},
  journal={Reliability Engineering \& System Safety},
  volume={207},
  pages={107365},
  year={2021},
  publisher={Elsevier}
}

@article{cheng2020surrogate,
  title={Surrogate-assisted global sensitivity analysis: an overview},
  author={Cheng, Kai and Lu, Zhenzhou and Ling, Chunyan and Zhou, Suting},
  journal={Structural and Multidisciplinary Optimization},
  volume={61},
  number={3},
  pages={1187--1213},
  year={2020},
  publisher={Springer}
}

@book{krishna2017digital,
  title={Digital signal processing algorithms: number theory, convolution, fast Fourier transforms, and applications},
  author={Krishna, Hari},
  year={2017},
  publisher={Routledge}
}

@article{han2022survey,
  title={A survey on vision transformer},
  author={Han, Kai and Wang, Yunhe and Chen, Hanting and Chen, Xinghao and Guo, Jianyuan and Liu, Zhenhua and Tang, Yehui and Xiao, An and Xu, Chunjing and Xu, Yixing and others},
  journal={IEEE transactions on pattern analysis and machine intelligence},
  volume={45},
  number={1},
  pages={87--110},
  year={2022},
  publisher={IEEE}
}

@article{zhang2018visual,
  title={Visual interpretability for deep learning: a survey},
  author={Zhang, Quan-shi and Zhu, Song-Chun},
  journal={Frontiers of Information Technology \& Electronic Engineering},
  volume={19},
  number={1},
  pages={27--39},
  year={2018},
  publisher={Springer}
}

@inproceedings{chakraborty2017interpretability,
  title={Interpretability of deep learning models: A survey of results},
  author={Chakraborty, Supriyo and Tomsett, Richard and Raghavendra, Ramya and Harborne, Daniel and Alzantot, Moustafa and Cerutti, Federico and Srivastava, Mani and Preece, Alun and Julier, Simon and Rao, Raghuveer M and others},
  booktitle={2017 IEEE smartworld, ubiquitous intelligence \& computing, advanced \& trusted computed, scalable computing \& communications, cloud \& big data computing, Internet of people and smart city innovation (smartworld/SCALCOM/UIC/ATC/CBDcom/IOP/SCI)},
  pages={1--6},
  year={2017},
  organization={IEEE}
}

@article{ferrante2025towards,
  title={Towards neural foundation models for vision: Aligning EEG, MEG, and fMRI representations for decoding, encoding, and modality conversion},
  author={Ferrante, Matteo and Boccato, Tommaso and Rashkov, Grigorii and Toschi, Nicola},
  journal={Information Fusion},
  pages={103650},
  year={2025},
  publisher={Elsevier}
}

@article{jia2019quantum,
  title={Quantum neural network states: A brief review of methods and applications},
  author={Jia, Zhih-Ahn and Yi, Biao and Zhai, Rui and Wu, Yu-Chun and Guo, Guang-Can and Guo, Guo-Ping},
  journal={Advanced Quantum Technologies},
  volume={2},
  number={7-8},
  pages={1800077},
  year={2019},
  publisher={Wiley Online Library}
}

@article{reh2023optimizing,
  title={Optimizing design choices for neural quantum states},
  author={Reh, Moritz and Schmitt, Markus and G{\"a}rttner, Martin},
  journal={Physical Review B},
  volume={107},
  number={19},
  pages={195115},
  year={2023},
  publisher={APS}
}

@article{cuomo2022scientific,
  title={Scientific machine learning through physics--informed neural networks: Where we are and what’s next},
  author={Cuomo, Salvatore and Di Cola, Vincenzo Schiano and Giampaolo, Fabio and Rozza, Gianluigi and Raissi, Maziar and Piccialli, Francesco},
  journal={Journal of Scientific Computing},
  volume={92},
  number={3},
  pages={88},
  year={2022},
  publisher={Springer}
}

@article{cai2021physics,
  title={Physics-informed neural networks (PINNs) for fluid mechanics: A review},
  author={Cai, Shengze and Mao, Zhiping and Wang, Zhicheng and Yin, Minglang and Karniadakis, George Em},
  journal={Acta Mechanica Sinica},
  volume={37},
  number={12},
  pages={1727--1738},
  year={2021},
  publisher={Springer}
}

@article{o2015introduction,
  title={An introduction to convolutional neural networks},
  author={O'shea, Keiron and Nash, Ryan},
  journal={arXiv preprint arXiv:1511.08458},
  year={2015}
}

@misc{matlab2025deeplearning,
  title        = {Deep Learning Toolbox},
  author       = {{MathWorks}},
  howpublished = {\url{https://www.mathworks.com/products/deep-learning.html}},
  year         = {2025},
  note         = {Accessed: 2025-12-09}
}

@article{d2020two,
  title={Two-dimensional topological quantum walks in the momentum space of structured light},
  author={D’Errico, Alessio and Cardano, Filippo and Maffei, Maria and Dauphin, Alexandre and Barboza, Raouf and Esposito, Chiara and Piccirillo, Bruno and Lewenstein, Maciej and Massignan, Pietro and Marrucci, Lorenzo},
  journal={Optica},
  volume={7},
  number={2},
  pages={108--114},
  year={2020},
  publisher={Optical Society of America}
}

@article{doersch2016tutorial,
  title={Tutorial on variational autoencoders},
  author={Doersch, Carl},
  journal={arXiv preprint arXiv:1606.05908},
  year={2016}
}

@article{schouten2005plasmon,
  title={Plasmon-assisted two-slit transmission: Young’s experiment revisited},
  author={Schouten, HF and Kuzmin, N and Dubois, G and Visser, TD and Gbur, G and Alkemade, PFA and Blok, H and Hooft, GW’t and Lenstra, D and Eliel, ER},
  journal={Physical Review Letters},
  volume={94},
  number={5},
  pages={053901},
  year={2005},
  publisher={APS}
}

@article{sawant2013non,
  title={Non-classical paths in interference experiments},
  author={Sawant, Rahul and Samuel, Joseph and Sinha, Aninda and Sinha, Supurna and Sinha, Urbasi},
  journal={arXiv preprint arXiv:1308.2022},
  year={2013}
}

@article{maaten2008visualizing,
  title={Visualizing data using t-SNE},
  author={Maaten, Laurens van der and Hinton, Geoffrey},
  journal={Journal of machine learning research},
  volume={9},
  number={Nov},
  pages={2579--2605},
  year={2008}
}

@article{bakurov2022structural,
  title={Structural similarity index (SSIM) revisited: A data-driven approach},
  author={Bakurov, Illya and Buzzelli, Marco and Schettini, Raimondo and Castelli, Mauro and Vanneschi, Leonardo},
  journal={Expert Systems with Applications},
  volume={189},
  pages={116087},
  year={2022},
  publisher={Elsevier}
}

@article{chazal2021introduction,
  title={An introduction to topological data analysis: fundamental and practical aspects for data scientists},
  author={Chazal, Fr{\'e}d{\'e}ric and Michel, Bertrand},
  journal={Frontiers in artificial intelligence},
  volume={4},
  pages={667963},
  year={2021},
  publisher={Frontiers Media SA}
}

@inproceedings{zhang2017degree,
  title={Degree centrality, betweenness centrality, and closeness centrality in social network},
  author={Zhang, Junlong and Luo, Yu},
  booktitle={2017 2nd international conference on modelling, simulation and applied mathematics (MSAM2017)},
  pages={300--303},
  year={2017},
  organization={Atlantis press}
}

@article{tarnowski2019measuring,
  title={Measuring topology from dynamics by obtaining the Chern number from a linking number},
  author={Tarnowski, Matthias and {\"U}nal, F Nur and Fl{\"a}schner, Nick and Rem, Benno S and Eckardt, Andr{\'e} and Sengstock, Klaus and Weitenberg, Christof},
  journal={Nature communications},
  volume={10},
  number={1},
  pages={1728},
  year={2019},
  publisher={Nature Publishing Group UK London}
}

@article{schroter2020observation,
  title={Observation and control of maximal Chern numbers in a chiral topological semimetal},
  author={Schr{\"o}ter, Niels BM and Stolz, Samuel and Manna, Kaustuv and De Juan, Fernando and Vergniory, Maia G and Krieger, Jonas A and Pei, Ding and Schmitt, Thorsten and Dudin, Pavel and Kim, Timur K and others},
  journal={Science},
  volume={369},
  number={6500},
  pages={179--183},
  year={2020},
  publisher={American Association for the Advancement of Science}
}

@article{pratt1973atomic,
  title={Atomic photoelectric effect above 10 keV},
  author={Pratt, RH and Ron, Akiva and Tseng, HK},
  journal={Reviews of Modern physics},
  volume={45},
  number={2},
  pages={273},
  year={1973},
  publisher={APS}
}

@article{paul1980einstein,
  title={The Einstein Podolsky Rosen Paradox and Local Hidden—Variables Theories},
  author={Paul, H},
  journal={Fortschritte der Physik},
  volume={28},
  number={12},
  pages={633--657},
  year={1980},
  publisher={Wiley Online Library}
}

@article{walborn2011revealing,
  title={Revealing hidden einstein-podolsky-rosen nonlocality},
  author={Walborn, SP and Salles, A and Gomes, RM and Toscano, F and Souto Ribeiro, PH},
  journal={Physical review letters},
  volume={106},
  number={13},
  pages={130402},
  year={2011},
  publisher={APS}
}

@article{aspect1982experimental,
  title={Experimental realization of Einstein-Podolsky-Rosen-Bohm Gedankenexperiment: a new violation of Bell's inequalities},
  author={Aspect, Alain and Grangier, Philippe and Roger, G{\'e}rard},
  journal={Physical review letters},
  volume={49},
  number={2},
  pages={91},
  year={1982},
  publisher={APS}
}

@article{healey1997nonlocality,
  title={Nonlocality and the Aharonov-Bohm effect},
  author={Healey, Richard},
  journal={Philosophy of Science},
  volume={64},
  number={1},
  pages={18--41},
  year={1997},
  publisher={Cambridge University Press}
}

@incollection{peres1997quantum,
  title={Quantum nonlocality and inseparability},
  author={Peres, Asher},
  booktitle={New Developments on Fundamental Problems in Quantum Physics},
  pages={301--310},
  year={1997},
  publisher={Springer}
}

@article{horodecki2009quantum,
  title={Quantum entanglement},
  author={Horodecki, Ryszard and Horodecki, Pawe{\l} and Horodecki, Micha{\l} and Horodecki, Karol},
  journal={Reviews of modern physics},
  volume={81},
  number={2},
  pages={865--942},
  year={2009},
  publisher={APS}
}

@article{amselem2009experimental,
  title={Experimental four-qubit bound entanglement},
  author={Amselem, Elias and Bourennane, Mohamed},
  journal={Nature Physics},
  volume={5},
  number={10},
  pages={748--752},
  year={2009},
  publisher={Nature Publishing Group UK London}
}

@article{herzog1995complementarity,
  title={Complementarity and the quantum eraser},
  author={Herzog, Thomas J and Kwiat, Paul G and Weinfurter, Harald and Zeilinger, Anton},
  journal={Physical Review Letters},
  volume={75},
  number={17},
  pages={3034},
  year={1995},
  publisher={APS}
}

@article{merzbacher2002early,
  title={The early history of quantum tunneling},
  author={Merzbacher, Eugen},
  journal={Physics Today},
  volume={55},
  number={8},
  pages={44--49},
  year={2002},
  publisher={AIP Publishing}
}

@article{shante1971introduction,
  title={An introduction to percolation theory},
  author={Shante, Vinod KS and Kirkpatrick, Scott},
  journal={Advances in Physics},
  volume={20},
  number={85},
  pages={325--357},
  year={1971},
  publisher={Taylor \& Francis}
}

@article{essam1980percolation,
  title={Percolation theory},
  author={Essam, John W},
  journal={Reports on progress in physics},
  volume={43},
  number={7},
  pages={833},
  year={1980},
  publisher={IOP Publishing}
}

@article{snyder1976goos,
  title={Goos-h{\"a}nchen shift},
  author={Snyder, Allan W and Love, John D},
  journal={Applied optics},
  volume={15},
  number={1},
  pages={236--238},
  year={1976},
  publisher={OSA}
}

@article{ayatollah2023can,
  title={Can the double-slit experiment distinguish between quantum interpretations?},
  author={Ayatollah Rafsanjani, Ali and Kazemi, MohammadJavad and Bahrampour, Alireza and Golshani, Mehdi},
  journal={Communications Physics},
  volume={6},
  number={1},
  pages={195},
  year={2023},
  publisher={Nature Publishing Group UK London}
}

@article{kazemi2023detection,
  title={Detection statistics in a double-double-slit experiment},
  author={Kazemi, MohammadJavad and Hosseinzadeh, Vahid},
  journal={Physical Review A},
  volume={107},
  number={1},
  pages={012223},
  year={2023},
  publisher={APS}
}

@article{koushki2023study,
  title={A study of the effects of Gaussian distribution of coherence length of source on the diffraction of partial temporal coherence beam from multi slits: Theory and simulation},
  author={Koushki, E and Alavi, SA},
  journal={Results in Optics},
  volume={13},
  pages={100546},
  year={2023},
  publisher={Elsevier}
}

@article{redding2011spatial,
  title={Spatial coherence of random laser emission},
  author={Redding, Brandon and Choma, Michael A and Cao, Hui},
  journal={Optics letters},
  volume={36},
  number={17},
  pages={3404--3406},
  year={2011},
  publisher={Optical Society of America}
}

@article{erkmen2010ghost,
  title={Ghost imaging: from quantum to classical to computational},
  author={Erkmen, Baris I and Shapiro, Jeffrey H},
  journal={Advances in Optics and Photonics},
  volume={2},
  number={4},
  pages={405--450},
  year={2010},
  publisher={Optical Society of America}
}

@article{padgett2017introduction,
  title={An introduction to ghost imaging: quantum and classical},
  author={Padgett, Miles J and Boyd, Robert W},
  journal={Philosophical Transactions of the Royal Society A: Mathematical, Physical and Engineering Sciences},
  volume={375},
  number={2099},
  pages={20160233},
  year={2017},
  publisher={The Royal Society Publishing}
}

@article{jafferis2022stringy,
  title={Stringy ER= EPR},
  author={Jafferis, Daniel Louis and Schneider, Elliot},
  journal={Journal of High Energy Physics},
  volume={2022},
  number={10},
  pages={1--86},
  year={2022},
  publisher={Springer}
}

@article{susskind2016er,
  title={ER= EPR, GHZ, and the consistency of quantum measurements},
  author={Susskind, Leonard},
  journal={Fortschritte der Physik},
  volume={64},
  number={1},
  pages={72--83},
  year={2016},
  publisher={Wiley Online Library}
}

@article{dai2020testing,
  title={Testing the ER= EPR conjecture},
  author={Dai, De-Chang and Minic, Djordje and Stojkovic, Dejan and Fu, Changbo},
  journal={Physical Review D},
  volume={102},
  number={6},
  pages={066004},
  year={2020},
  publisher={APS}
}

@book{rubinstein2016simulation,
  title={Simulation and the Monte Carlo method},
  author={Rubinstein, Reuven Y and Kroese, Dirk P},
  year={2016},
  publisher={John Wiley \& Sons}
}

@inproceedings{ho2024online,
  title={Online transit entanglement routing in quantum networks: Architecture design and optimization},
  author={Ho, Wan-Ting and Fang, Shing-Yan and Hsieh, Wei-Chia and Cheni, Li-Feng and Kuo, Jian-Jhih and Tsai, Ming-Jer},
  booktitle={GLOBECOM 2024-2024 IEEE Global Communications Conference},
  pages={4497--4502},
  year={2024},
  organization={IEEE}
}

@article{huang2025decoherence,
  title={Decoherence-Aware Entangling and Swapping Strategy Optimization for Entanglement Routing in Quantum Networks},
  author={Huang, Shao-Min and Cheng, Cheng-Yang and Chien, Ming-Huang and Kuo, Jian-Jhih and Wang, Chih-Yu},
  journal={arXiv preprint arXiv:2510.14912},
  year={2025}
}

@article{teng2013near,
  title={Near-field interference of slit doublet},
  author={Teng, Shuyun and Li, Furui and Wang, Junhong and Zhang, Wei},
  journal={Journal of the Optical Society of America A},
  volume={30},
  number={11},
  pages={2273--2279},
  year={2013},
  publisher={Optical Society of America}
}

\end{document}